\documentclass[prl,twocolumn,showpacs,citeautoscript,superscriptaddress,longbibliography]{revtex4-1}

\usepackage{pdfpages}

\usepackage{amsmath}
\usepackage{amssymb}
\usepackage{amsfonts}
\usepackage{graphicx,graphics}
\usepackage{bm}  
\usepackage{float}

\newcommand{\abs}[1]{\lvert #1 \rvert}
\newcommand{\expect}[1]{\langle #1\rangle}

\newcommand{\br}{\mathbf{r}}

\begin{document}

\title{Theory of light emission from quantum noise in plasmonic contacts:
  above-threshold emission from higher-order electron-plasmon scattering}
\author{Kristen Kaasbjerg}
\email{cosby@fys.ku.dk}
\affiliation{Department of Condensed Matter Physics, Weizmann Institute of
  Science, Rehovot 76100, Israel}
\affiliation{School of Chemistry, The Sackler Faculty of Exact Sciences, Tel Aviv
  University, Tel Aviv 69978, Israel}
\author{Abraham Nitzan}
\affiliation{School of Chemistry, The Sackler Faculty of Exact Sciences, Tel Aviv
  University, Tel Aviv 69978, Israel}
\date{\today}

\begin{abstract}
  We develop a theoretical framework for the description of light emission from
  plasmonic contacts based on the nonequilibrium Green function formalism. Our
  theory establishes a fundamental link between the finite-frequency quantum
  noise and AC conductance of the contact and the light emission. Calculating
  the quantum noise to higher orders in the electron-plasmon interaction, we
  identify a plasmon-induced electron-electron interaction as the source of
  experimentally observed above-threshold light emission from biased STM
  contacts. Our findings provide important insight into the effect of
  interactions on the light emission from atomic-scale contacts.
\end{abstract}

\maketitle

\emph{Introduction.}---Atomic-size contacts~\cite{Ruitenbeek:AtomicSized},
realized, e.g., in scanning tunneling microscopy (STM), provide a unique
platform to study fundamental quantum transport phenomena such as, e.g.,
conductance quantization~\cite{Norskov:Quantized}, suppression of shot
noise~\cite{Ruitenbeek:QuantumSuppression}, and vibrational inelastic effects on
the conductance and shot
noise~\cite{Vieira:Onset,Ruitenbeek:Detection}. Recently, light emitted from STM
contacts due to inelastic electron scattering off localized plasmons was used as
a probe of the shot noise at optical
frequencies~\cite{Berndt:OpticalProbe}. Apart from standard emission due to
one-electron scattering processes with photon energy $\hbar\omega < eV$, the
observation of above-threshold emission with $\hbar\omega > eV$ indicates that
interaction effects on the noise are
probed~\cite{Berndt:AtomContact,Berndt:OpticalProbe}. Emission from atomic-scale
contacts is thus well-suited for studying the fundamental properties of
high-frequency quantum noise.

The role of quantum noise as excitation source of electromagnetic fields is
known from theoretical work on mesoscopic
conductors~\cite{Loosen:On,Kouwenhoven:DQD,Imry:Detection,Schomerus:Counting}.
The emission is related to the positive frequency part,
$\mathcal{S}(\omega > 0)$, of the \emph{asymmetric} quantum shot noise
\begin{equation}
  \label{eq:S}
  \mathcal{S}(\omega) = \int\! dt \, e^{-i \omega t} \expect{\delta I(0) \delta I(t)},
\end{equation}
where $\delta I = I - \expect{I}$ and $I$ is the current operator, whereas the
absorption is connected to the negative frequency part,
$\mathcal{S}(\omega<0)$. The study of finite-frequency noise and the associated
emission has been attracting attention in
atomic-scale~\cite{Berndt:AtomContact,Berndt:OpticalProbe},
molecular~\cite{Hou:Generation,Berndt:LightEmission,Nitzan:Opto,Brandbyge:LightEmission,Schull:Electroluminescence}
and mesoscopic
conductors~\cite{Jin:Experimental,Aharony:NoiseSpectra,Schoelkopt:Introduction,Deblock:Emission,Blatter:Statistics,Esteve:Bright,Komnik:Anderson,Safi:Onechannel,Portier:DBShotNoise,Belzig:NonGaussian,Reulet:Emission},
and may shed light on fundamental issues in molecular
optoelectronics~\cite{Nitzan:Opto} and quantum
plasmonics~\cite{Baumberg:Revealing,Novotny:Electrical,Kim:QuantumPlasmonics} as
well as the origin of above-threshold emission. For the latter, different
interactions have been
addressed~\cite{Johansson:TwoElectron,Nazarov:QuantumTunneling,Berndt:HotElectron,Belzig:NonGaussian}.
However, a complete picture and a systematic treatment are lacking.

\begin{figure}[!b]
  \includegraphics[width=0.9\linewidth]{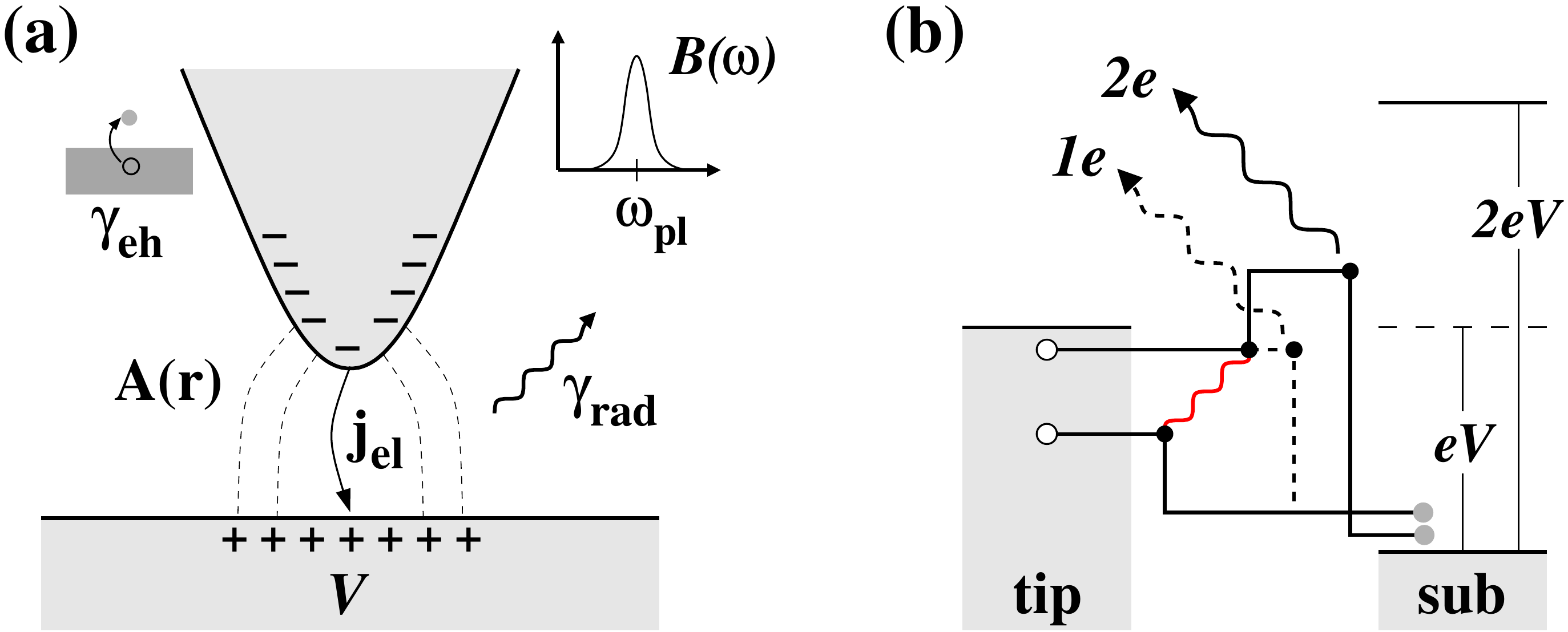}
  \caption{(Color online) (a) Plasmonic STM contact. The inset illustrates the
    spectrum of the localized surface-plasmon polarition (LSP) supported by the
    contact. (b) Schematic illustration of the one and two-electron scattering
    processes responsible for the $1e$ ($\hbar\omega<eV$; dashed lines) and $2e$
    ($eV<\hbar\omega<2eV$; full lines) light emission.}
\label{fig:overview}
\end{figure}

Here, we develop a microscopic theory for plasmonic light emission in
atomic-scale contacts based on the Keldysh nonequilibrium Green function (GF)
formalism~\cite{Rammer:QuantumField,Jauho}. Our approach allows for a systematic calculation of the
light emission to higher orders in the interaction between the tunnel current
and localized surface-plasmon polaritions (LSPs) supported by the contact. LSPs
are instrumental to light
emission~\cite{Muhlschlegel:Role,Apell:Theory,Johansson:Inelastic,Novotny:Electrical}
and serve as a direct probe of the quantum noise~\cite{Berndt:OpticalProbe} due
to their radiative nature and the enhanced electron-plasmon (el-pl) interaction
which results from the strong local fields associated with them (see
Fig.~\ref{fig:overview}(a))~\cite{Cuevas:Field}. We here establish the link
between the quantum noise and plasmonic light within the Keldysh GF formalism
and furthermore discuss the relation to the AC conductance. Studying a generic
contact model, we find that the experimentally observed above-threshold
emission~\cite{Berndt:AtomContact,Berndt:OpticalProbe} stems from higher-order
contributions to the quantum noise associated with the plasmon-induced
two-electron scattering process illustrated in Fig.~\ref{fig:overview}(b).

\emph{Theory.}---We consider a single radiative LSP mode with frequency
$\omega_\text{pl}$ (see Fig.~\ref{fig:overview}(a)) represented by the quantized
vector potential field
\begin{equation}
  \mathbf{A}(\br) = \bm{\xi}_\text{pl}(\br) 
      \sqrt{\frac{\hbar}{2 \Omega \epsilon_0 \omega_\text{pl}}}
      \left( a^\dagger + a \right) ,
\end{equation}
where $\Omega$ is a quantization volume and $\bm{\xi}_\text{pl}$ is a mode
vector describing the spatial distribution of the
field~\cite{Johansson:Light}. The interaction between the LSP and the tunnel
current is given by the standard coupling term
\begin{equation}
  \label{eq:H_elpl}
  H_\text{el-pl} = \int \! d\br \, \mathbf{j}_\text{el}(\br)\cdot\mathbf{A}(\br) ,
\end{equation}
where $\mathbf{j}_\text{el}=\mathbf{j}_\text{el}^\nabla +
\mathbf{j}_\text{el}^A$ is the electronic current density, and
$\mathbf{j}_\text{el}^\nabla$ and $\mathbf{j}_\text{el}^A$ the paramagnetic and
diamagnetic components, respectively.

The excitation dynamics of the LSP and the emission is encoded in the LSP GF
$D(\tau,\tau')=-i \expect{T_c A(\tau)A(\tau')}$ where $A=a + a^\dagger$ and
$T_c$ is the time-ordering operator on the Keldysh contour. The retarded and
lesser components of the GF are given by their respective Dyson and Keldysh
equations,
\begin{equation}
  D^r(\omega)  = d^r_0(\omega) + d_0^r(\omega) 
                 \Pi^r(\omega) D^r(\omega) ,
\end{equation}
and
\begin{equation}
  D^<(\omega)  = D^r(\omega) \Pi^<(\omega) D^a(\omega) ,
\end{equation}
where $d_0^r(\omega)= \tfrac{ 2\omega_\text{pl} } {(\omega + i0^+)^2 -
  \omega_\text{pl}^2}$ is the \emph{bare} GF and $\Pi=\Pi_\text{el} +
\Pi_\text{rad} + \Pi_\text{eh}$ is the \emph{irreducible} self-energy which
accounts for the interaction with the tunneling current ($\Pi_\text{el}$) as
well as radiative decay ($\Pi_\text{rad}$) and decay into bulk electron-hole
pair excitation ($\Pi_\text{eh}$). The latter two are modeled with
phenomenological damping parameters $\gamma_\text{rad/eh}$ with self-energies
$\Pi_\text{rad/eh}^r(\omega)= -i \gamma_\text{rad/eh} \mathrm{sgn} (\omega) /2$
and $\Pi_\text{rad/eh}^{>/<}(\omega)= -i \gamma_\text{rad/eh}
\abs{n_B(\mp\omega)}$, where $n_B$ is the Bose-Einstein distribution function,
and give rise to a broadened LSP, $D_0^r(\omega) = \tfrac{2 \omega_\text{pl}}
{\omega^2 - \omega_\text{pl}^2 - i \omega_\text{pl} \gamma_0}$, with width
$\gamma_0 = \gamma_\text{rad} + \gamma_\text{eh}$.

The excitation, damping and broadening of the LSP due to the el-pl
interaction~\eqref{eq:H_elpl} are governed by the el-pl self-energy,
$\Pi_\text{el} = \Pi^\nabla + \Pi^A$. To lowest order in the interaction, the
two contributions are given by~\cite{Mahan}
\begin{equation}
  \label{eq:Pi}
  \Pi^\nabla(\tau,\tau') = -i \expect{T_c \delta j(\tau) \delta j(\tau')}_0 ,
\end{equation}
and
\begin{equation}
  \Pi^A(\tau,\tau') = \delta(\tau-\tau') \expect{\rho(\tau)}_0 ,
\end{equation}
respectively, where $\delta j = j - \expect{j}_0$,
\begin{equation}
  \label{eq:j}
  j = \sqrt{\frac{\hbar}{2 \Omega \epsilon_0 \omega_\text{pl}}}
     \int \! d\br \, \bm{\xi}_\text{pl}(\br) \cdot 
     \mathbf{j}_\text{el}^\nabla(\br)
\end{equation}
is the projection of the current operator onto the LSP mode vector,
\begin{equation}
  \rho = -\frac{e^2}{m_e} 
     \frac{\hbar}{2 \Omega \epsilon_0 \omega_\text{pl}}
     \int \! d\br \, \bm{\xi}_\text{pl}(\br) \cdot 
     \bm{\xi}_\text{pl}(\br) \Psi^\dagger(\br) \Psi(\br) 
\end{equation}
is the mode-weighted electron-density operator and $\expect{\cdot}_0$ is the
expectation value in the absence of the el-pl interaction~\eqref{eq:H_elpl}.
The diamagnetic self-energy does not affect the LSP dynamics as it only depends
on the static charge density $\expect{\rho}_0$, and is here neglected.

From a perturbative expansion of the LSP GF in the paramagnetic interaction, we
find that the \emph{exact} paramagnetic self-energy is given
by~\cite{supplemental}
\begin{equation}
  \label{Pi:exact}
  \Pi^\nabla(\tau,\tau') = -i S^\text{irr}(\tau,\tau')  ,
\end{equation}
valid to all orders in the el-pl interaction. Here, $S^\text{irr}$ is the
\emph{irreducible} part of the current-current correlation function
$S(\tau,\tau') = \expect{T_c \delta j(\tau) \delta j(\tau')}$~\cite{footnote3}
and the expectation value $\expect{\cdot}$ is with respect to the Hamiltonian
that includes the el-pl interaction~\eqref{eq:H_elpl}.

Focusing on the case where the el-pl coupling~\eqref{eq:j} is proportional to
the current operator $I$, the lesser and retarded components of the correlation
function $S$ are given by the quantum noise in Eq.~\eqref{eq:S} and the response
function $\,\mathcal{K}^r(t-t') = -i \Theta(t-t') \expect{[I(t),I(t') ]}$,
respectively. From the important formal result~\eqref{Pi:exact}, it then follows
that the lesser $\Pi^{\nabla,<}=-iS_\text{irr}^<$ and retarded
$\Pi^{\nabla,r}=-iS_\text{irr}^r$ self-energies are directly related to these
quantities, thus connecting the damping of the LSP to the dissipative real part
of the AC conductance $\mathcal{G}(\omega) = i \mathcal{K}^r(\omega)/\omega$ of
the contact. These quantities are connected via~\cite{footnote1}
\begin{equation}
  -2\omega \text{Re}\mathcal{G} = 2 \text{Im} \mathcal{K}^r(\omega) = \mathcal{S}(\omega) - \mathcal{S}(-\omega) ,
\end{equation}
which may be regarded as a \emph{nonequilibrium} fluctuation-dissipation
relation. This demonstrates a fundamental connection between quantum noise and
AC conductance previously discussed in
Refs.~\onlinecite{Loosen:On,Levinson:Quantum}.

To connect the quantum noise to the light emission, we consider the radiative
decay of the LSP into a reservoir of far-field modes, $H_\text{far-field} =
\sum_\lambda \hbar \omega_\lambda a_\lambda^\dagger
a_\lambda^{\phantom\dagger}$, with the associated exchange rate per unit
frequency given by
\begin{equation}
  \label{eq:emission}
  \Gamma_\text{rad}(\omega) = \Pi_\text{rad}^<(\omega) D^>(\omega) 
                             - \Pi_\text{rad}^>(\omega) D^<(\omega).
\end{equation}
Here, the two terms account for absorption and emission, respectively, in
agreement with Eq.~\eqref{eq:S}~\cite{footnote1}. The emitted light is thus
governed by $D^<(\omega) = - B(\omega)
\tfrac{\Pi^<(\omega)}{2\text{Im}\Pi^r(\omega)} \overset{T=0, \atop \omega>0}{=}
- B(\omega) \tfrac{\Pi_\text{el}^<(\omega)}{2\text{Im}\Pi^r(\omega)}$, where
$B=-2\text{Im}D^r$ is the LSP spectral function (see
Fig.~\ref{fig:overview}(a)), showing that it resembles the LSP spectrum and is
driven by $\Pi_\text{el}^<$. With the above, we have established the link
between plasmonic light emission and the quantum noise and AC conductance.

\emph{Generic model and results.}---With the formal theory established, we go on
to study the light emission and finite-frequency noise in a generic model for an
atomic-scale STM contact consisting of a spin-degenerate electronic
state/conduction channel coupled to bulk lead reservoirs of the STM tip and
substrate,
\begin{equation}
  \label{eq:H}
  H  = \varepsilon_0 \sum_\sigma d_\sigma^\dagger d^{\phantom\dagger}_\sigma 
        + \sum_\alpha H_\alpha + H_T 
        + H_\text{el-pl} 
        + \hbar\omega_\text{pl} a^\dagger a .
\end{equation}
Here, $\varepsilon_0=0$ is the energy of the electronic level and $H_\alpha =
\sum_k \varepsilon_{k,\alpha} c_{k\alpha}^\dagger
c_{k\alpha}^{\phantom\dagger}$, $\alpha=\{\text{tip},\text{sub}\}$ is the
Hamiltonian of the reservoirs with chemical potentials $\mu_\text{tip/sub}=\pm
V/2$. The coupling to the reservoirs, $H_T = \sum_{\alpha, k} t_k^\alpha
(c_{k\alpha}^ \dagger d + \text{h.c.}  )$, leads to tunnel broadenings
$\Gamma_\alpha = 2\pi \rho_{\alpha} \abs{t_{\alpha}}^2$ which define tunneling
($\Gamma_\text{sub} \ll \Gamma_\text{tip}$) and contact ($\Gamma_\text{sub} \sim
\Gamma_\text{tip}$) regimes. For $\Gamma = \Gamma_\text{tip} +
\Gamma_\text{sub}\gg eV, k_\text{B}T$, the DC conductance is given by
$\mathcal{G}=G_0 T$ where $G_0=2e^2/h$ and $T =
4\Gamma_\text{tip}\Gamma_\text{sub} / \Gamma^2$ is the transmission coefficient.

The el-pl interaction takes the form $H_\text{el-pl} = \sum_\alpha M_\alpha
I_\alpha (a^\dagger + a)$, where $I_\alpha= i \sum_k (t_\alpha c_{k\alpha}^
\dagger d - \text{h.c.})$ is the paramagnetic current operator at reservoir
$\alpha$ and $M_\alpha = \tfrac{e}{\hbar} \sqrt{\tfrac{\hbar}{2 \Omega
    \epsilon_0 \omega_\text{pl}}} l_\alpha$ is a dimensionless coupling constant
with $l_\alpha$ a characteristic length scale for the
interaction~\cite{Scalapino:TheCriteria}. To describe the experimental light
emission~\cite{Berndt:AtomContact,Berndt:OpticalProbe}, we take the el-pl
coupling to be given by $\abs{M_\text{tip/sub}}=M$ with
$M_\text{tip}=-M_\text{sub}$, which implies that the LSP couples to the total
current $I_\text{tot} = I_\text{tip} - I_\text{sub}$ through the
contact~\cite{footnote2}.  The paramagnetic self-energy can now be written as a
sum over lead-lead components, $\Pi^\nabla=\sum_{\alpha\beta}
\Pi_{\alpha\beta}^\nabla$, where
\begin{align}
  \label{eq:Pi_ab}
  \Pi_{\alpha\beta}^\nabla(\tau,\tau') 
  = -i M_\alpha M_\beta S_{\alpha\beta}^\text{irr}(\tau,\tau') ,
\end{align}
with $S_{\alpha\beta}(\tau,\tau') = \expect{T_c \delta I_\alpha(\tau) \delta
  I_\beta(\tau')}$ and $\delta I_\alpha = I_\alpha - \expect{I_\alpha}$. With
the above-mentioned assumption for the coupling constant, we have
$\Pi^\nabla(\tau,\tau') = -i M^2 S^\text{irr}(\tau,\tau')$ where $S(\tau,\tau')
= \expect{T_c \delta I_\text{tot}(\tau) \delta I_\text{tot}(\tau')}$. 
\begin{figure}[!t]
  \includegraphics[width=0.9\linewidth]{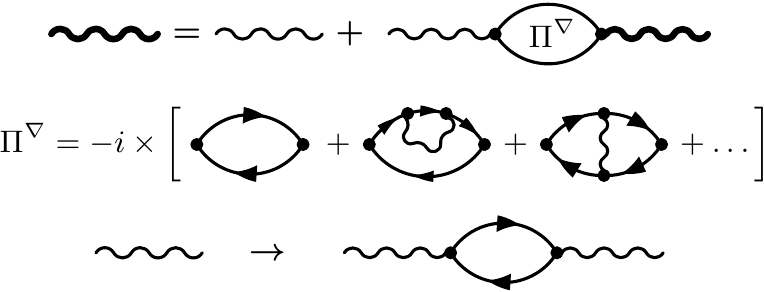}
  \caption{(Color online) (top) Dyson equation for the LSP GF where $\Pi^\nabla$
    denotes the paramagnetic el-pl self-energy. (center) Perturbation expansion
    of the self-energy showing the diagrams up to 4'th order in the el-pl
    interaction. (bottom) The 6'th order diagrams responsible for the $2e$
    emission can be obtained by replacing the broadened LSP GF with its 2'nd
    order correction (a plasmon excited by electron tunneling) in the 4'th order
    diagrams. Feynman dictionary: $\bullet$: el-pl interaction; solid lines:
    electronic contact GF $G_{ij}(\tau,\tau') = -i \expect{T_c c_i(\tau)
      c_j^\dagger(\tau')}_0, \quad i,j = \text{tip}, \text{sub}, d$; thin wiggly
    lines: broadened LSP GF $D_0 (\tau,\tau')=-i \expect{T_c
      A(\tau)A(\tau')}_0$.}
\label{fig:feynman}
\end{figure}

In the following, we proceed with a perturbative calculation of the
\emph{irreducible} el-pl self-energy illustrated in terms of Feynman diagrams in
Fig.~\ref{fig:feynman} (see Ref.~\cite{supplemental} for details). We focus on
the regime $\Gamma \gg \hbar\omega_\text{pl}, eV, k_\text{B}T$ and take
$k_\text{B}T=0$, corresponding to the experimentally relevant situation
$k_\text{B}T \ll \hbar\omega_\text{pl}$ where the current is the only excitation
source for the LSP.

\begin{figure}[!t]
  \centering
  \includegraphics[width=1.0\linewidth]{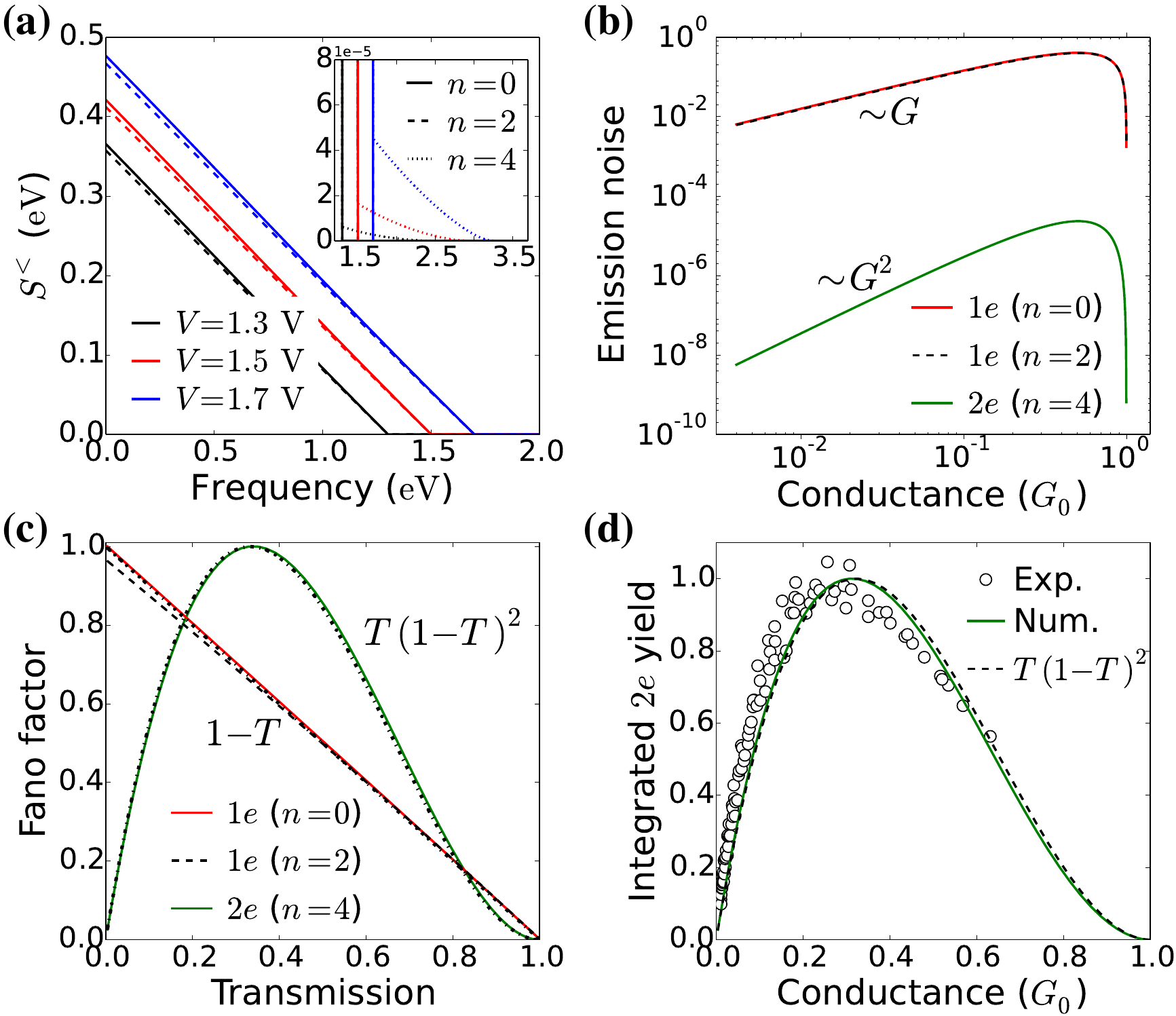}
  \caption{(Color online) (a) Noise spectrum to different orders in the el-pl
    interaction ($n=0$: solid, $n=2$: dashed, and $n=4$: dotted lines) for a
    contact with $T\sim 0.2$. The inset shows a zoom of the noise spectrum at
    $\omega\gtrsim eV$. (b) and (c) Integrated $1e$ and $2e$ emission noise and
    Fano factors at $V=1.6$~V vs conductance and transmission, respectively. The
    dash-dotted lines in (c) show the indicated analytic functions. (d)
    Integrated $2e$ emission yield vs conductance at $V=1.6$~V. The circles show
    the experimental $2e$ yield from Ref.~\cite{Berndt:OpticalProbe}. In (c) and (d),
    the Fano factors and yields have been normalized to unity at their maximum
    value (the $1e$ Fano factors have been normalized with the $n=0$
    maximum). Parameters: $\omega_\text{pl} = 1.5$~eV, $\gamma_0=0.2$~eV,
    $M=0.1$, $\Gamma_\text{tip}=10$~eV.}
\label{fig:noise}
\end{figure}

Figure~\ref{fig:noise}(a) shows the numerically calculated (\emph{irreducible})
emission noise to different orders in the el-pl interaction (the order of the
corresponding self-energy is $n+2$) for a contact with $T\sim0.2$ and LSP
parameters resembling the experiment~\cite{Berndt:OpticalProbe}.
The lowest-order ($n=0$) \emph{noninteracting} quantum noise (full lines in
Fig.~\ref{fig:noise}(a)) is given by the \emph{bare} bubble diagram. In the
limit $\Gamma \gg eV, \hbar\omega, k_\text{B}T$, we find, in agreement with
previous works~\cite{Kouwenhoven:DQD,Buttiker:ShotNoise}, that the
\emph{noninteracting} noise spectrum is given by~\cite{footnote4}
\begin{align}
  \label{eq:Snonint}
  S_0^<(\omega) & \approx  \frac{4\times 2}{2\pi} \big[
       T (1 - T) \left[ H(\omega + eV) + H(\omega - eV) \right] 
       \bigg. \nonumber \\ & \quad \bigg.
  + 2 T^2 H(\omega) \big] ,
\end{align}
where $H(x) = x n_B(x)$. At $k_\text{B}T=0$, the emission part simplifies to
$S_0^<(\omega>0) \sim T(1-T) \Theta(eV - \omega) (eV-\omega)$, and is hence
suppressed at perfect transmission and cut off at $\omega=eV$, i.e. in this
order only emission with $\omega < eV$ ($1e$) via the one-electron scattering
process illustrated in Fig.~\ref{fig:overview}(b) (dashed lines) is included.

The lowest-order corrections ($n=2$)---a bubble with a ``plasmon-dressed''
contact GF and one with a vertex correction---give rise to a reduction of the
emission noise (dashed lines in Fig.~\ref{fig:noise}(a) and are cut off at
$\omega=eV$ like the \emph{noninteracting} noise.

The above-threshold emission is contained in the 4'th order ($n=4$) quantum
noise. At $k_\text{B}T=0$, it originates from the coherent two-electron
scattering process illustrated in Fig.~\ref{fig:overview}(b), which corresponds
to a plasmon-induced electron-electron (el-el) interaction. To identify the
associated self-energy diagrams we employ the \emph{optical theorem} stating
that the imaginary part of a self-energy diagram is given by the sum of squared
scattering amplitudes from possible ``on-shell'' cuts of the
self-energy~\cite{Zee}. The relevant self-energy diagrams are therefore
identified as the two $n=2$ diagrams with the damped LSP GF replaced by its
lowest-order correction (see bottom Fig.~\ref{fig:feynman}). The two diagrams
give rise to the $\omega > eV$ component ($2e$) of the noise shown in the inset
of Fig.~\ref{fig:noise}(a) (dotted lines) which is cut off at $\omega=2eV$. The
cutoff stems from the above-mentioned scattering process, where an initial
emission process exciting the plasmon [wiggly (red) line in
Fig.~\ref{fig:overview}(b)] is followed by an absorption process generating a
``hot electron'' which can emit at above-threshold energies $eV< \omega<2eV$. At
finite temperature, ``hot electrons'' can also be generated from thermally
excited plasmons. However, at $k_\text{B}T\ll \hbar\omega_\text{pl}$ this
process is negligible.

To further analyze the quantum noise, we show in Figure~\ref{fig:noise}(b) the
integrated $1e$ ($0<\omega<eV$) and $2e$ ($eV<\omega<2 eV$)
components~\cite{Berndt:AtomContact,Berndt:OpticalProbe} of the emission noise
at $V=1.6$~V as a function of the conductance. In the tunneling regime,
$\mathcal{G} \ll G_0$, the noise components are found to scale with the
conductance as $S_{1e}^< \sim \mathcal{G}$ and $S_{2e}^< \sim \mathcal{G}^2$,
respectively, emphasizing that they involve one and two-electron scattering
processes. Moreover, at perfect transmission, $\mathcal{G} \sim G_0$, both the
$1e$ and $2e$ noise are suppressed. For the $1e$ noise, the $n=2$ corrections do
not change this qualitatively. We emphasize that the suppression of emission
noise at $T=1$ only holds in the considered large-$\Gamma$
limit~\cite{supplemental}.

The fact that the $2e$ noise shows a $\mathcal{G}^2$ dependence in the tunneling
regime and is suppressed at contact, suggests that it scales as the square of
the prefactor in the \emph{noninteracting} noise, i.e. $S_{2e}^< \sim T^2 (1 -
T)^2$. To test this hypothesis, we inspect the finite-frequency Fano factor
$F(\omega) = \mathcal{S}(\omega) / eI$ which should then scale with the
transmission coefficient as
\begin{equation}
  \label{eq:F_scaling}
  F_{1e} \sim 1 - T \quad \text{and} \quad  F_{2e} \sim T(1 - T)^2 ,
\end{equation}
with the maximum of $F_{2e}$ occurring at $T=1/3$. The integrated Fano factors
shown in Fig.~\ref{fig:noise}(c) are in excellent agreement with our
expectations, confirming the anticipated scaling of the $2e$ emission noise. The
ratio of the $1e$ and $2e$ emission noise thus scales with the coupling constant
and transmission coefficient as $S_{2e}^< / S_{1e}^< \sim M^4 T(1-T)$.

\begin{figure}[!t]
  \centering
  \includegraphics[width=1.0\linewidth]{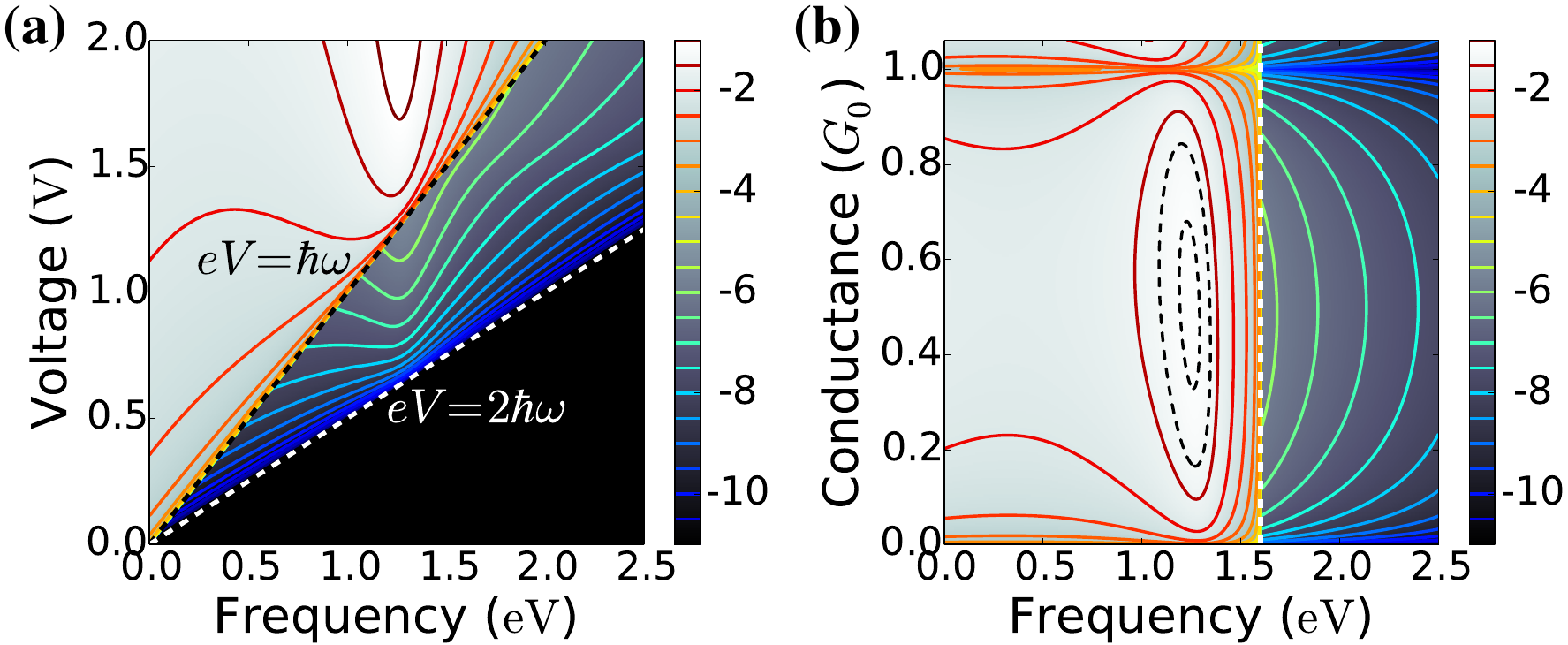}
  \caption{(Color online) (a) Emission spectrum vs applied voltage for a contact
    with $T\sim 0.2$. (b) Emission spectrum vs transmission coefficient at
    $V=1.6$~V. The plots shows the emission rate
    $\Gamma_\text{rad}(\omega)\propto -\text{Im} D^<(\omega)$ in units of
    $\gamma_\text{rad}$ on a logarithmic scale. Parameters: $\omega_\text{pl} =
    1.5$~eV, $\gamma_0=0.2$~eV, $M=0.1$, $\Gamma_\text{tip}=10$~eV.}
\label{fig:emission}
\end{figure}
Next, we discuss the emission spectrum shown in Fig.~\ref{fig:emission} as a
function of bias voltage and conductance. As expected, the emission which
resembles the LSP spectrum has a dominant $1e$ component which is driven by the
\emph{noninteracting} quantum noise, and a weaker $2e$ component driven by the
higher-order quantum noise. Due to the predicted $T$ dependence of the quantum
noise, both the $1e$ and $2e$ emission peak at $\mathcal{G} \sim 0.5$~$G_0$ and
are strongly reduced at $\mathcal{G} \sim G_0$. This counter intuitive behavior
at perfect transmission where the current is maximized and one naively would
expect the same for the emission, is a unique fingerprint of the quantum noise
origin.

The tunneling-induced damping of the LSP associated with the dissipative part of
the AC conductance gives rise to an additional spectral broadening
$\gamma_\text{el} = -2 \text{Im} \Pi^{\nabla,r}$. In the large $\Gamma$-limit
and to lowest order in the el-pl interaction, $\text{Re} \mathcal{G}(\omega) =
G_0 T$ and $\gamma_\text{el} = 8 \omega / \pi M^2 T$. Contrary to the emission
noise, it does not vanish at $T=1$ and is independent of the bias voltage. Due
to the nondissipative part of the AC conductance (real part of the self-energy)
the LSP resonance redshifts ($\sim 0.1$ eV) with increasing conductance in
Fig.~\ref{fig:emission}(b). Similar spectral features have been observed in
subnanometer plasmonic contacts~\cite{Baumberg:Revealing,Dionne:Observation},
though the importance of tunneling versus other mechanisms is
unclear~\cite{Gao:QMStudy,Aizpurua:Bridging,Bozhevolnyi:Generalized,Borisov:Robust,Borisov:Quantum}.
Our findings indicate that electron tunneling plays a nonnegligible role in the
quantum regime.

Finally, in Fig.~\ref{fig:noise}(d) we show the calculated $2e$ emission yield,
defined as emission per current, as a function of conductance together with the
experimental $2e$ photon yield from Ref.~\cite{Berndt:OpticalProbe}. Compared to
the Fano factor in Fig.~\ref{fig:noise}(c), the above-mentioned spectral changes
result in a slight left shift of the curve for the yield. The agreement with the
experimental $2e$ yield is very good, indicating that we have identified the
mechanism responsible for $2e$ emission. At $\mathcal{G}\sim G_0$, however, the
experimental yields do not show complete suppression (see
Ref.~\cite{Berndt:OpticalProbe}). This discrepancy can be due to experimental
factors such as: (i) imperfect or additional transmission
channels~\cite{Berndt:OpticalProbe}, and/or (ii) changes in the LSP mode and
el-pl coupling as the tip-substrate distance is
reduced~\cite{Berndt:Electromagnetic}.

\emph{Summary.}---To summarize, we have presented a framework based on the
Keldysh GF formalism for the description of light emission from plasmonic
contacts and established the connection between the quantum noise and AC
conductance of the contact and the light emission. Studying a generic contact
model, we have identified a plasmon-induced el-el interaction associated with
the higher-order quantum noise as the mechanisms behind the experimentally
observed above-threshold
emission~\cite{Berndt:AtomContact,Berndt:OpticalProbe}. Our approach, which can
be generalized to more complex situations, paves the way for a better
understanding of the effect of interactions on light emission and quantum noise
in atomic-scale and molecular contacts.

\begin{acknowledgments}
  We would like to thank W.~Belzig, T.~Novotn{\'y}, M.~Galperin,
  N.~A. Mortensen, J.~Paaske and M.~Brandbyge for fruitful discussions,
  R. Berndt for providing us with the original data from
  Ref.~\onlinecite{Berndt:OpticalProbe} and M.~H.~Fischer for comments on the
  manuscript. This research was supported by the Israel Science Foundation, the
  Israel-US Binational Science Foundation (grant No. 2011509), and the European
  Science Council (FP7/ERC grant No. 226628). AN thanks the Physics Department
  at FUB and SFB 658 for hospitality during the time this work was completed. KK
  acknowledges support from the Villum and Carlsberg Foundations.
\end{acknowledgments}

\bibliography{paper.bbl}

\begin{thebibliography}{57}%
\makeatletter
\providecommand \@ifxundefined [1]{%
 \@ifx{#1\undefined}
}%
\providecommand \@ifnum [1]{%
 \ifnum #1\expandafter \@firstoftwo
 \else \expandafter \@secondoftwo
 \fi
}%
\providecommand \@ifx [1]{%
 \ifx #1\expandafter \@firstoftwo
 \else \expandafter \@secondoftwo
 \fi
}%
\providecommand \natexlab [1]{#1}%
\providecommand \enquote  [1]{``#1''}%
\providecommand \bibnamefont  [1]{#1}%
\providecommand \bibfnamefont [1]{#1}%
\providecommand \citenamefont [1]{#1}%
\providecommand \href@noop [0]{\@secondoftwo}%
\providecommand \href [0]{\begingroup \@sanitize@url \@href}%
\providecommand \@href[1]{\@@startlink{#1}\@@href}%
\providecommand \@@href[1]{\endgroup#1\@@endlink}%
\providecommand \@sanitize@url [0]{\catcode `\\12\catcode `\$12\catcode
  `\&12\catcode `\#12\catcode `\^12\catcode `\_12\catcode `\%12\relax}%
\providecommand \@@startlink[1]{}%
\providecommand \@@endlink[0]{}%
\providecommand \url  [0]{\begingroup\@sanitize@url \@url }%
\providecommand \@url [1]{\endgroup\@href {#1}{\urlprefix }}%
\providecommand \urlprefix  [0]{URL }%
\providecommand \Eprint [0]{\href }%
\providecommand \doibase [0]{http://dx.doi.org/}%
\providecommand \selectlanguage [0]{\@gobble}%
\providecommand \bibinfo  [0]{\@secondoftwo}%
\providecommand \bibfield  [0]{\@secondoftwo}%
\providecommand \translation [1]{[#1]}%
\providecommand \BibitemOpen [0]{}%
\providecommand \bibitemStop [0]{}%
\providecommand \bibitemNoStop [0]{.\EOS\space}%
\providecommand \EOS [0]{\spacefactor3000\relax}%
\providecommand \BibitemShut  [1]{\csname bibitem#1\endcsname}%
\let\auto@bib@innerbib\@empty
\bibitem [{\citenamefont {Agra{\"\i}t}\ \emph {et~al.}(2003)\citenamefont
  {Agra{\"\i}t}, \citenamefont {Yeyati},\ and\ \citenamefont {van
  Ruitenbeek}}]{Ruitenbeek:AtomicSized}%
  \BibitemOpen
  \bibfield  {author} {\bibinfo {author} {\bibfnamefont {N.}~\bibnamefont
  {Agra{\"\i}t}}, \bibinfo {author} {\bibfnamefont {A.~Levy}\ \bibnamefont
  {Yeyati}}, \ and\ \bibinfo {author} {\bibfnamefont {J.~M.}\ \bibnamefont {van
  Ruitenbeek}},\ }\bibfield  {title} {\enquote {\bibinfo {title} {Quantum
  properties of atomic-sized conductors},}\ }\href@noop {} {\bibfield
  {journal} {\bibinfo  {journal} {Physics Reports}\ }\textbf {\bibinfo {volume}
  {377}},\ \bibinfo {pages} {81} (\bibinfo {year} {2003})}\BibitemShut
  {NoStop}%
\bibitem [{\citenamefont {Olesen}\ \emph {et~al.}(1994)\citenamefont {Olesen},
  \citenamefont {Laegsgaard}, \citenamefont {Stensgaard}, \citenamefont
  {Besenbacher}, \citenamefont {Schi{\o}tz}, \citenamefont {Stoltze},
  \citenamefont {Jacobsen},\ and\ \citenamefont
  {N{\o}rskov}}]{Norskov:Quantized}%
  \BibitemOpen
  \bibfield  {author} {\bibinfo {author} {\bibfnamefont {L.}~\bibnamefont
  {Olesen}}, \bibinfo {author} {\bibfnamefont {E.}~\bibnamefont {Laegsgaard}},
  \bibinfo {author} {\bibfnamefont {I.}~\bibnamefont {Stensgaard}}, \bibinfo
  {author} {\bibfnamefont {F.}~\bibnamefont {Besenbacher}}, \bibinfo {author}
  {\bibfnamefont {J.}~\bibnamefont {Schi{\o}tz}}, \bibinfo {author}
  {\bibfnamefont {P.}~\bibnamefont {Stoltze}}, \bibinfo {author} {\bibfnamefont
  {K.~W.}\ \bibnamefont {Jacobsen}}, \ and\ \bibinfo {author} {\bibfnamefont
  {J.~K.}\ \bibnamefont {N{\o}rskov}},\ }\bibfield  {title} {\enquote {\bibinfo
  {title} {Quantized conductance in an atom-sized point contact},}\ }\href@noop
  {} {\bibfield  {journal} {\bibinfo  {journal} {Phys. Rev. Lett.}\ }\textbf
  {\bibinfo {volume} {72}},\ \bibinfo {pages} {2251} (\bibinfo {year}
  {1994})}\BibitemShut {NoStop}%
\bibitem [{\citenamefont {van~den Brom}\ and\ \citenamefont {van
  Ruitenbeek}(1999)}]{Ruitenbeek:QuantumSuppression}%
  \BibitemOpen
  \bibfield  {author} {\bibinfo {author} {\bibfnamefont {H.~E.}\ \bibnamefont
  {van~den Brom}}\ and\ \bibinfo {author} {\bibfnamefont {J.~M.}\ \bibnamefont
  {van Ruitenbeek}},\ }\bibfield  {title} {\enquote {\bibinfo {title} {Quantum
  suppression of shot noise in atom-size metallic contacts},}\ }\href@noop {}
  {\bibfield  {journal} {\bibinfo  {journal} {Phys. Rev. Lett.}\ }\textbf
  {\bibinfo {volume} {82}},\ \bibinfo {pages} {1526} (\bibinfo {year}
  {1999})}\BibitemShut {NoStop}%
\bibitem [{\citenamefont {Agra{\"\i}t}\ \emph {et~al.}(2002)\citenamefont
  {Agra{\"\i}t}, \citenamefont {Untiedt}, \citenamefont {Rubio-Bollinger},\
  and\ \citenamefont {Vieira}}]{Vieira:Onset}%
  \BibitemOpen
  \bibfield  {author} {\bibinfo {author} {\bibfnamefont {N.}~\bibnamefont
  {Agra{\"\i}t}}, \bibinfo {author} {\bibfnamefont {C.}~\bibnamefont
  {Untiedt}}, \bibinfo {author} {\bibfnamefont {G.}~\bibnamefont
  {Rubio-Bollinger}}, \ and\ \bibinfo {author} {\bibfnamefont {S.}~\bibnamefont
  {Vieira}},\ }\bibfield  {title} {\enquote {\bibinfo {title} {Electron
  transport and phonons in atomic wires},}\ }\href@noop {} {\bibfield
  {journal} {\bibinfo  {journal} {Phys. Rev. Lett.}\ }\textbf {\bibinfo
  {volume} {88}},\ \bibinfo {pages} {216803} (\bibinfo {year}
  {2002})}\BibitemShut {NoStop}%
\bibitem [{\citenamefont {Kumar}\ \emph {et~al.}(2012)\citenamefont {Kumar},
  \citenamefont {Avriller}, \citenamefont {Yeyati},\ and\ \citenamefont {{van
  Ruitenbeek}}}]{Ruitenbeek:Detection}%
  \BibitemOpen
  \bibfield  {author} {\bibinfo {author} {\bibfnamefont {M.}~\bibnamefont
  {Kumar}}, \bibinfo {author} {\bibfnamefont {R.}~\bibnamefont {Avriller}},
  \bibinfo {author} {\bibfnamefont {A.~L.}\ \bibnamefont {Yeyati}}, \ and\
  \bibinfo {author} {\bibfnamefont {J.~M.}\ \bibnamefont {{van Ruitenbeek}}},\
  }\bibfield  {title} {\enquote {\bibinfo {title} {Detection of vibration-mode
  scattering in electronic shot noise},}\ }\href@noop {} {\bibfield  {journal}
  {\bibinfo  {journal} {Phys. Rev. Lett.}\ }\textbf {\bibinfo {volume} {108}},\
  \bibinfo {pages} {146602} (\bibinfo {year} {2012})}\BibitemShut {NoStop}%
\bibitem [{\citenamefont {Schneider}\ \emph {et~al.}(2010)\citenamefont
  {Schneider}, \citenamefont {Schull},\ and\ \citenamefont
  {Berndt}}]{Berndt:OpticalProbe}%
  \BibitemOpen
  \bibfield  {author} {\bibinfo {author} {\bibfnamefont {N.~L.}\ \bibnamefont
  {Schneider}}, \bibinfo {author} {\bibfnamefont {G.}~\bibnamefont {Schull}}, \
  and\ \bibinfo {author} {\bibfnamefont {R.}~\bibnamefont {Berndt}},\
  }\bibfield  {title} {\enquote {\bibinfo {title} {Optical probe of quantum
  shot-noise reduction at a single-atom contact},}\ }\href@noop {} {\bibfield
  {journal} {\bibinfo  {journal} {Phys. Rev. Lett.}\ }\textbf {\bibinfo
  {volume} {105}},\ \bibinfo {pages} {026601} (\bibinfo {year}
  {2010})}\BibitemShut {NoStop}%
\bibitem [{\citenamefont {Schull}\ \emph {et~al.}(2009)\citenamefont {Schull},
  \citenamefont {N{\'e}el}, \citenamefont {Johansson},\ and\ \citenamefont
  {Berndt}}]{Berndt:AtomContact}%
  \BibitemOpen
  \bibfield  {author} {\bibinfo {author} {\bibfnamefont {G.}~\bibnamefont
  {Schull}}, \bibinfo {author} {\bibfnamefont {N.}~\bibnamefont {N{\'e}el}},
  \bibinfo {author} {\bibfnamefont {P.}~\bibnamefont {Johansson}}, \ and\
  \bibinfo {author} {\bibfnamefont {R.}~\bibnamefont {Berndt}},\ }\bibfield
  {title} {\enquote {\bibinfo {title} {Electron-plasmon and electron-electron
  interactions at a single atom contact},}\ }\href@noop {} {\bibfield
  {journal} {\bibinfo  {journal} {Phys. Rev. Lett.}\ }\textbf {\bibinfo
  {volume} {102}},\ \bibinfo {pages} {057401} (\bibinfo {year}
  {2009})}\BibitemShut {NoStop}%
\bibitem [{\citenamefont {Lesovik}\ and\ \citenamefont
  {Loosen}(1997)}]{Loosen:On}%
  \BibitemOpen
  \bibfield  {author} {\bibinfo {author} {\bibfnamefont {G.~B.}\ \bibnamefont
  {Lesovik}}\ and\ \bibinfo {author} {\bibfnamefont {R.}~\bibnamefont
  {Loosen}},\ }\bibfield  {title} {\enquote {\bibinfo {title} {On the detection
  of finite-frequency current fluctuations},}\ }\href@noop {} {\bibfield
  {journal} {\bibinfo  {journal} {Pis'ma Zh. Eksp. Teor. Fiz.}\ }\textbf
  {\bibinfo {volume} {65}},\ \bibinfo {pages} {280} (\bibinfo {year} {1997})},\
  \bibinfo {note} {[JETP Lett. 65, 295 (1997)]}\BibitemShut {NoStop}%
\bibitem [{\citenamefont {Aguado}\ and\ \citenamefont
  {Kouwenhoven}(2000)}]{Kouwenhoven:DQD}%
  \BibitemOpen
  \bibfield  {author} {\bibinfo {author} {\bibfnamefont {R.}~\bibnamefont
  {Aguado}}\ and\ \bibinfo {author} {\bibfnamefont {L.~P.}\ \bibnamefont
  {Kouwenhoven}},\ }\bibfield  {title} {\enquote {\bibinfo {title} {Double
  quantum dots as detectors of high-frequency quantum noise in mesoscopic
  conductors},}\ }\href@noop {} {\bibfield  {journal} {\bibinfo  {journal}
  {Phys. Rev. Lett.}\ }\textbf {\bibinfo {volume} {84}},\ \bibinfo {pages}
  {1986} (\bibinfo {year} {2000})}\BibitemShut {NoStop}%
\bibitem [{\citenamefont {Gavish}\ \emph {et~al.}(2000)\citenamefont {Gavish},
  \citenamefont {Levinson},\ and\ \citenamefont {Imry}}]{Imry:Detection}%
  \BibitemOpen
  \bibfield  {author} {\bibinfo {author} {\bibfnamefont {U.}~\bibnamefont
  {Gavish}}, \bibinfo {author} {\bibfnamefont {Y.}~\bibnamefont {Levinson}}, \
  and\ \bibinfo {author} {\bibfnamefont {Y.}~\bibnamefont {Imry}},\ }\bibfield
  {title} {\enquote {\bibinfo {title} {Detection of quantum noise},}\
  }\href@noop {} {\bibfield  {journal} {\bibinfo  {journal} {Phys. Rev. B}\
  }\textbf {\bibinfo {volume} {62}},\ \bibinfo {pages} {10637} (\bibinfo {year}
  {2000})}\BibitemShut {NoStop}%
\bibitem [{\citenamefont {Beenakker}\ and\ \citenamefont
  {Schomerus}(2001)}]{Schomerus:Counting}%
  \BibitemOpen
  \bibfield  {author} {\bibinfo {author} {\bibfnamefont {C.~W.~J.}\
  \bibnamefont {Beenakker}}\ and\ \bibinfo {author} {\bibfnamefont
  {H.}~\bibnamefont {Schomerus}},\ }\bibfield  {title} {\enquote {\bibinfo
  {title} {Counting statistics of photons produced by electronic shot noise},}\
  }\href@noop {} {\bibfield  {journal} {\bibinfo  {journal} {Phys. Rev. Lett.}\
  }\textbf {\bibinfo {volume} {86}},\ \bibinfo {pages} {700} (\bibinfo {year}
  {2001})}\BibitemShut {NoStop}%
\bibitem [{\citenamefont {Dong}\ \emph {et~al.}(2010)\citenamefont {Dong},
  \citenamefont {Zhang}, \citenamefont {Gao}, \citenamefont {Luo},
  \citenamefont {Zhang}, \citenamefont {Chen}, \citenamefont {Zhang},
  \citenamefont {Tao}, \citenamefont {Zhang}, \citenamefont {Yang},\ and\
  \citenamefont {Hou}}]{Hou:Generation}%
  \BibitemOpen
  \bibfield  {author} {\bibinfo {author} {\bibfnamefont {Z.~C.}\ \bibnamefont
  {Dong}}, \bibinfo {author} {\bibfnamefont {X.~L.}\ \bibnamefont {Zhang}},
  \bibinfo {author} {\bibfnamefont {H.~Y.}\ \bibnamefont {Gao}}, \bibinfo
  {author} {\bibfnamefont {Y.}~\bibnamefont {Luo}}, \bibinfo {author}
  {\bibfnamefont {C.}~\bibnamefont {Zhang}}, \bibinfo {author} {\bibfnamefont
  {L.~G.}\ \bibnamefont {Chen}}, \bibinfo {author} {\bibfnamefont
  {R.}~\bibnamefont {Zhang}}, \bibinfo {author} {\bibfnamefont
  {X.}~\bibnamefont {Tao}}, \bibinfo {author} {\bibfnamefont {Y.}~\bibnamefont
  {Zhang}}, \bibinfo {author} {\bibfnamefont {J.~L.}\ \bibnamefont {Yang}}, \
  and\ \bibinfo {author} {\bibfnamefont {J.~G.}\ \bibnamefont {Hou}},\
  }\bibfield  {title} {\enquote {\bibinfo {title} {Generation of molecular hot
  electroluminescence by resonant nanocavity plasmons},}\ }\href@noop {}
  {\bibfield  {journal} {\bibinfo  {journal} {Nature Phot.}\ }\textbf {\bibinfo
  {volume} {4}},\ \bibinfo {pages} {50} (\bibinfo {year} {2010})}\BibitemShut
  {NoStop}%
\bibitem [{\citenamefont {Schneider}\ \emph {et~al.}(2012)\citenamefont
  {Schneider}, \citenamefont {L{\"u}}, \citenamefont {Brandbyge},\ and\
  \citenamefont {Berndt}}]{Berndt:LightEmission}%
  \BibitemOpen
  \bibfield  {author} {\bibinfo {author} {\bibfnamefont {N.~L.}\ \bibnamefont
  {Schneider}}, \bibinfo {author} {\bibfnamefont {J.~T.}\ \bibnamefont
  {L{\"u}}}, \bibinfo {author} {\bibfnamefont {M.}~\bibnamefont {Brandbyge}}, \
  and\ \bibinfo {author} {\bibfnamefont {R.}~\bibnamefont {Berndt}},\
  }\bibfield  {title} {\enquote {\bibinfo {title} {Light emission probing
  quantum shot noise and charge fluctuations at a biased molecular junction},}\
  }\href@noop {} {\bibfield  {journal} {\bibinfo  {journal} {Phys. Rev. Lett.}\
  }\textbf {\bibinfo {volume} {109}},\ \bibinfo {pages} {186601} (\bibinfo
  {year} {2012})}\BibitemShut {NoStop}%
\bibitem [{\citenamefont {Galperin}\ and\ \citenamefont
  {Nitzan}(2012)}]{Nitzan:Opto}%
  \BibitemOpen
  \bibfield  {author} {\bibinfo {author} {\bibfnamefont {M.}~\bibnamefont
  {Galperin}}\ and\ \bibinfo {author} {\bibfnamefont {A.}~\bibnamefont
  {Nitzan}},\ }\bibfield  {title} {\enquote {\bibinfo {title} {Molecular
  optoelectronics: the interaction of molecular conduction junctions with
  light},}\ }\href@noop {} {\bibfield  {journal} {\bibinfo  {journal} {Phys.
  Chem. Chem. Phys.}\ }\textbf {\bibinfo {volume} {14}},\ \bibinfo {pages}
  {9421} (\bibinfo {year} {2012})}\BibitemShut {NoStop}%
\bibitem [{\citenamefont {L{\"u}}\ \emph {et~al.}(2013)\citenamefont {L{\"u}},
  \citenamefont {Christensen},\ and\ \citenamefont
  {Brandbyge}}]{Brandbyge:LightEmission}%
  \BibitemOpen
  \bibfield  {author} {\bibinfo {author} {\bibfnamefont {J.-T.}\ \bibnamefont
  {L{\"u}}}, \bibinfo {author} {\bibfnamefont {R.~Bjerregaard}\ \bibnamefont
  {Christensen}}, \ and\ \bibinfo {author} {\bibfnamefont {M.}~\bibnamefont
  {Brandbyge}},\ }\bibfield  {title} {\enquote {\bibinfo {title} {Light
  emission and finite-frequency shot noise in molecular junctions: {From}
  tunneling to contact},}\ }\href@noop {} {\bibfield  {journal} {\bibinfo
  {journal} {Phys. Rev. B}\ }\textbf {\bibinfo {volume} {88}},\ \bibinfo
  {pages} {045413} (\bibinfo {year} {2013})}\BibitemShut {NoStop}%
\bibitem [{\citenamefont {Reecht}\ \emph {et~al.}(2014)\citenamefont {Reecht},
  \citenamefont {Scheurer}, \citenamefont {Speisser}, \citenamefont {Dappe},
  \citenamefont {Mathevet},\ and\ \citenamefont
  {Schull}}]{Schull:Electroluminescence}%
  \BibitemOpen
  \bibfield  {author} {\bibinfo {author} {\bibfnamefont {G.}~\bibnamefont
  {Reecht}}, \bibinfo {author} {\bibfnamefont {F.}~\bibnamefont {Scheurer}},
  \bibinfo {author} {\bibfnamefont {V.}~\bibnamefont {Speisser}}, \bibinfo
  {author} {\bibfnamefont {Y.~J.}\ \bibnamefont {Dappe}}, \bibinfo {author}
  {\bibfnamefont {F.}~\bibnamefont {Mathevet}}, \ and\ \bibinfo {author}
  {\bibfnamefont {G.}~\bibnamefont {Schull}},\ }\bibfield  {title} {\enquote
  {\bibinfo {title} {Electroluminescence of a polythiophene molecular wire
  suspended between a metallic surface and the tip of a scanning tunneling
  microscope},}\ }\href@noop {} {\bibfield  {journal} {\bibinfo  {journal}
  {Phys. Rev. Lett.}\ }\textbf {\bibinfo {volume} {112}},\ \bibinfo {pages}
  {047403} (\bibinfo {year} {2014})}\BibitemShut {NoStop}%
\bibitem [{\citenamefont {Zakka-Bajjani}\ \emph {et~al.}(2007)\citenamefont
  {Zakka-Bajjani}, \citenamefont {S{\'e}gala}, \citenamefont {Portier},
  \citenamefont {Roche}, \citenamefont {Glattli}, \citenamefont {Cavanna},\
  and\ \citenamefont {Jin}}]{Jin:Experimental}%
  \BibitemOpen
  \bibfield  {author} {\bibinfo {author} {\bibfnamefont {E.}~\bibnamefont
  {Zakka-Bajjani}}, \bibinfo {author} {\bibfnamefont {J.}~\bibnamefont
  {S{\'e}gala}}, \bibinfo {author} {\bibfnamefont {F.}~\bibnamefont {Portier}},
  \bibinfo {author} {\bibfnamefont {P.}~\bibnamefont {Roche}}, \bibinfo
  {author} {\bibfnamefont {D.~C.}\ \bibnamefont {Glattli}}, \bibinfo {author}
  {\bibfnamefont {A.}~\bibnamefont {Cavanna}}, \ and\ \bibinfo {author}
  {\bibfnamefont {Y.}~\bibnamefont {Jin}},\ }\bibfield  {title} {\enquote
  {\bibinfo {title} {Experimental test of the high-frequency quantum shot noise
  theory in a quantum point contact},}\ }\href@noop {} {\bibfield  {journal}
  {\bibinfo  {journal} {Phys. Rev. Lett.}\ }\textbf {\bibinfo {volume} {99}},\
  \bibinfo {pages} {236803} (\bibinfo {year} {2007})}\BibitemShut {NoStop}%
\bibitem [{\citenamefont {Rothstein}\ \emph {et~al.}(2009)\citenamefont
  {Rothstein}, \citenamefont {{Entin-Wohlman}},\ and\ \citenamefont
  {Aharony}}]{Aharony:NoiseSpectra}%
  \BibitemOpen
  \bibfield  {author} {\bibinfo {author} {\bibfnamefont {E.~A.}\ \bibnamefont
  {Rothstein}}, \bibinfo {author} {\bibfnamefont {O.}~\bibnamefont
  {{Entin-Wohlman}}}, \ and\ \bibinfo {author} {\bibfnamefont {A.}~\bibnamefont
  {Aharony}},\ }\bibfield  {title} {\enquote {\bibinfo {title} {Noise spectra
  of a quantum dot},}\ }\href@noop {} {\bibfield  {journal} {\bibinfo
  {journal} {Phys. Rev. B}\ }\textbf {\bibinfo {volume} {79}},\ \bibinfo
  {pages} {075307} (\bibinfo {year} {2009})}\BibitemShut {NoStop}%
\bibitem [{\citenamefont {Clerk}\ \emph {et~al.}(2010)\citenamefont {Clerk},
  \citenamefont {Devoret}, \citenamefont {Girvin}, \citenamefont {Marquardt},\
  and\ \citenamefont {Schoelkopf}}]{Schoelkopt:Introduction}%
  \BibitemOpen
  \bibfield  {author} {\bibinfo {author} {\bibfnamefont {A.~A.}\ \bibnamefont
  {Clerk}}, \bibinfo {author} {\bibfnamefont {M.~H.}\ \bibnamefont {Devoret}},
  \bibinfo {author} {\bibfnamefont {S.~M.}\ \bibnamefont {Girvin}}, \bibinfo
  {author} {\bibfnamefont {F.}~\bibnamefont {Marquardt}}, \ and\ \bibinfo
  {author} {\bibfnamefont {R.~J.}\ \bibnamefont {Schoelkopf}},\ }\bibfield
  {title} {\enquote {\bibinfo {title} {Introduction to quantum noise,
  measurement, and amplification},}\ }\href@noop {} {\bibfield  {journal}
  {\bibinfo  {journal} {Rev. Mod. Phys.}\ }\textbf {\bibinfo {volume} {82}},\
  \bibinfo {pages} {1155} (\bibinfo {year} {2010})}\BibitemShut {NoStop}%
\bibitem [{\citenamefont {Basset}\ \emph {et~al.}(2010)\citenamefont {Basset},
  \citenamefont {Bouchiat},\ and\ \citenamefont {Deblock}}]{Deblock:Emission}%
  \BibitemOpen
  \bibfield  {author} {\bibinfo {author} {\bibfnamefont {J.}~\bibnamefont
  {Basset}}, \bibinfo {author} {\bibfnamefont {H.}~\bibnamefont {Bouchiat}}, \
  and\ \bibinfo {author} {\bibfnamefont {R.}~\bibnamefont {Deblock}},\
  }\bibfield  {title} {\enquote {\bibinfo {title} {Emission and absorption
  quantum noise measurement with an on-chip resonant circuit},}\ }\href@noop {}
  {\bibfield  {journal} {\bibinfo  {journal} {Phys. Rev. Lett.}\ }\textbf
  {\bibinfo {volume} {105}},\ \bibinfo {pages} {166801} (\bibinfo {year}
  {2010})}\BibitemShut {NoStop}%
\bibitem [{\citenamefont {Lebedev}\ \emph {et~al.}(2010)\citenamefont
  {Lebedev}, \citenamefont {Lesovik},\ and\ \citenamefont
  {Blatter}}]{Blatter:Statistics}%
  \BibitemOpen
  \bibfield  {author} {\bibinfo {author} {\bibfnamefont {A.~V.}\ \bibnamefont
  {Lebedev}}, \bibinfo {author} {\bibfnamefont {G.~B.}\ \bibnamefont
  {Lesovik}}, \ and\ \bibinfo {author} {\bibfnamefont {G.}~\bibnamefont
  {Blatter}},\ }\bibfield  {title} {\enquote {\bibinfo {title} {Statistics of
  radiation emitted from a quantum point contact},}\ }\href@noop {} {\bibfield
  {journal} {\bibinfo  {journal} {Phys. Rev. B}\ }\textbf {\bibinfo {volume}
  {81}},\ \bibinfo {pages} {155421} (\bibinfo {year} {2010})}\BibitemShut
  {NoStop}%
\bibitem [{\citenamefont {Hofheinz}\ \emph {et~al.}(2011)\citenamefont
  {Hofheinz}, \citenamefont {Portier}, \citenamefont {Baudouin}, \citenamefont
  {Joyez}, \citenamefont {Vion}, \citenamefont {Bertet}, \citenamefont
  {Roche},\ and\ \citenamefont {Esteve}}]{Esteve:Bright}%
  \BibitemOpen
  \bibfield  {author} {\bibinfo {author} {\bibfnamefont {M.}~\bibnamefont
  {Hofheinz}}, \bibinfo {author} {\bibfnamefont {F.}~\bibnamefont {Portier}},
  \bibinfo {author} {\bibfnamefont {Q.}~\bibnamefont {Baudouin}}, \bibinfo
  {author} {\bibfnamefont {P.}~\bibnamefont {Joyez}}, \bibinfo {author}
  {\bibfnamefont {D.}~\bibnamefont {Vion}}, \bibinfo {author} {\bibfnamefont
  {P.}~\bibnamefont {Bertet}}, \bibinfo {author} {\bibfnamefont
  {P.}~\bibnamefont {Roche}}, \ and\ \bibinfo {author} {\bibfnamefont
  {D.}~\bibnamefont {Esteve}},\ }\bibfield  {title} {\enquote {\bibinfo {title}
  {Bright side of the {Coulomb} blockade},}\ }\href@noop {} {\bibfield
  {journal} {\bibinfo  {journal} {prl}\ }\textbf {\bibinfo {volume} {106}},\
  \bibinfo {pages} {217005} (\bibinfo {year} {2011})}\BibitemShut {NoStop}%
\bibitem [{\citenamefont {Orth}\ \emph {et~al.}(2012)\citenamefont {Orth},
  \citenamefont {Urban},\ and\ \citenamefont {Komnik}}]{Komnik:Anderson}%
  \BibitemOpen
  \bibfield  {author} {\bibinfo {author} {\bibfnamefont {C.~P.}\ \bibnamefont
  {Orth}}, \bibinfo {author} {\bibfnamefont {D.~F.}\ \bibnamefont {Urban}}, \
  and\ \bibinfo {author} {\bibfnamefont {A.}~\bibnamefont {Komnik}},\
  }\bibfield  {title} {\enquote {\bibinfo {title} {Finite-frequency noise
  properties of the nonequilibrium {Anderson} impurity model},}\ }\href@noop {}
  {\bibfield  {journal} {\bibinfo  {journal} {Phys. Rev. B}\ }\textbf {\bibinfo
  {volume} {86}},\ \bibinfo {pages} {125324} (\bibinfo {year}
  {2012})}\BibitemShut {NoStop}%
\bibitem [{\citenamefont {Zamoum}\ \emph {et~al.}(2012)\citenamefont {Zamoum},
  \citenamefont {Cr\'epieux},\ and\ \citenamefont {Safi}}]{Safi:Onechannel}%
  \BibitemOpen
  \bibfield  {author} {\bibinfo {author} {\bibfnamefont {R.}~\bibnamefont
  {Zamoum}}, \bibinfo {author} {\bibfnamefont {A.}~\bibnamefont {Cr\'epieux}},
  \ and\ \bibinfo {author} {\bibfnamefont {I.}~\bibnamefont {Safi}},\
  }\bibfield  {title} {\enquote {\bibinfo {title} {One-channel conductor
  coupled to a quantum of resistance: {Exact} ac conductance and
  finite-frequency noise},}\ }\href@noop {} {\bibfield  {journal} {\bibinfo
  {journal} {Phys. Rev. B}\ }\textbf {\bibinfo {volume} {85}},\ \bibinfo
  {pages} {125421} (\bibinfo {year} {2012})}\BibitemShut {NoStop}%
\bibitem [{\citenamefont {Altimiras}\ \emph {et~al.}(2014)\citenamefont
  {Altimiras}, \citenamefont {Parlavecchio}, \citenamefont {Joyez},
  \citenamefont {Vion}, \citenamefont {Roche}, \citenamefont {Esteve},\ and\
  \citenamefont {Portier}}]{Portier:DBShotNoise}%
  \BibitemOpen
  \bibfield  {author} {\bibinfo {author} {\bibfnamefont {C.}~\bibnamefont
  {Altimiras}}, \bibinfo {author} {\bibfnamefont {O.}~\bibnamefont
  {Parlavecchio}}, \bibinfo {author} {\bibfnamefont {P.}~\bibnamefont {Joyez}},
  \bibinfo {author} {\bibfnamefont {D.}~\bibnamefont {Vion}}, \bibinfo {author}
  {\bibfnamefont {P.}~\bibnamefont {Roche}}, \bibinfo {author} {\bibfnamefont
  {D.}~\bibnamefont {Esteve}}, \ and\ \bibinfo {author} {\bibfnamefont
  {F.}~\bibnamefont {Portier}},\ }\bibfield  {title} {\enquote {\bibinfo
  {title} {Dynamical {Coulomb} blockade of shot noise},}\ }\href@noop {}
  {\bibfield  {journal} {\bibinfo  {journal} {Phys. Rev. Lett.}\ }\textbf
  {\bibinfo {volume} {112}},\ \bibinfo {pages} {236803} (\bibinfo {year}
  {2014})}\BibitemShut {NoStop}%
\bibitem [{\citenamefont {Xu}\ \emph {et~al.}(2014)\citenamefont {Xu},
  \citenamefont {Holmqvist},\ and\ \citenamefont
  {Belzig}}]{Belzig:NonGaussian}%
  \BibitemOpen
  \bibfield  {author} {\bibinfo {author} {\bibfnamefont {F.}~\bibnamefont
  {Xu}}, \bibinfo {author} {\bibfnamefont {C.}~\bibnamefont {Holmqvist}}, \
  and\ \bibinfo {author} {\bibfnamefont {W.}~\bibnamefont {Belzig}},\
  }\bibfield  {title} {\enquote {\bibinfo {title} {Overbias light emission due
  to higher-order quantum noise in a tunnel junction},}\ }\href@noop {}
  {\bibfield  {journal} {\bibinfo  {journal} {Phys. Rev. Lett.}\ }\textbf
  {\bibinfo {volume} {113}},\ \bibinfo {pages} {066801} (\bibinfo {year}
  {2014})}\BibitemShut {NoStop}%
\bibitem [{\citenamefont {Forgues}\ \emph {et~al.}(2014)\citenamefont
  {Forgues}, \citenamefont {Lupien},\ and\ \citenamefont
  {Reulet}}]{Reulet:Emission}%
  \BibitemOpen
  \bibfield  {author} {\bibinfo {author} {\bibfnamefont {J.-C.}\ \bibnamefont
  {Forgues}}, \bibinfo {author} {\bibfnamefont {C.}~\bibnamefont {Lupien}}, \
  and\ \bibinfo {author} {\bibfnamefont {B.}~\bibnamefont {Reulet}},\
  }\bibfield  {title} {\enquote {\bibinfo {title} {Emission of microwave photon
  pairs by a tunnel junction},}\ }\href@noop {} {\bibfield  {journal} {\bibinfo
   {journal} {Phys. Rev. Lett.}\ }\textbf {\bibinfo {volume} {113}},\ \bibinfo
  {pages} {043602} (\bibinfo {year} {2014})}\BibitemShut {NoStop}%
\bibitem [{\citenamefont {Savage}\ \emph {et~al.}(2012)\citenamefont {Savage},
  \citenamefont {Hawkeye}, \citenamefont {Esteban}, \citenamefont {Borisov},
  \citenamefont {Aizpurua},\ and\ \citenamefont
  {Baumberg}}]{Baumberg:Revealing}%
  \BibitemOpen
  \bibfield  {author} {\bibinfo {author} {\bibfnamefont {K.~J.}\ \bibnamefont
  {Savage}}, \bibinfo {author} {\bibfnamefont {M.~M.}\ \bibnamefont {Hawkeye}},
  \bibinfo {author} {\bibfnamefont {R.}~\bibnamefont {Esteban}}, \bibinfo
  {author} {\bibfnamefont {A.~G.}\ \bibnamefont {Borisov}}, \bibinfo {author}
  {\bibfnamefont {J.}~\bibnamefont {Aizpurua}}, \ and\ \bibinfo {author}
  {\bibfnamefont {J.~J.}\ \bibnamefont {Baumberg}},\ }\bibfield  {title}
  {\enquote {\bibinfo {title} {Revealing the quantum regime in tunnelling
  plasmonics},}\ }\href@noop {} {\bibfield  {journal} {\bibinfo  {journal}
  {Nature}\ }\textbf {\bibinfo {volume} {491}},\ \bibinfo {pages} {574}
  (\bibinfo {year} {2012})}\BibitemShut {NoStop}%
\bibitem [{\citenamefont {Bharadwaj}\ \emph {et~al.}(2011)\citenamefont
  {Bharadwaj}, \citenamefont {Bouhelier},\ and\ \citenamefont
  {Novotny}}]{Novotny:Electrical}%
  \BibitemOpen
  \bibfield  {author} {\bibinfo {author} {\bibfnamefont {P.}~\bibnamefont
  {Bharadwaj}}, \bibinfo {author} {\bibfnamefont {A.}~\bibnamefont
  {Bouhelier}}, \ and\ \bibinfo {author} {\bibfnamefont {L.}~\bibnamefont
  {Novotny}},\ }\bibfield  {title} {\enquote {\bibinfo {title} {Electrical
  excitation of surface plasmons},}\ }\href@noop {} {\bibfield  {journal}
  {\bibinfo  {journal} {Phys. Rev. Lett.}\ }\textbf {\bibinfo {volume} {106}},\
  \bibinfo {pages} {226802} (\bibinfo {year} {2011})}\BibitemShut {NoStop}%
\bibitem [{\citenamefont {Tame}\ \emph {et~al.}(2013)\citenamefont {Tame},
  \citenamefont {McEnery}, \citenamefont {\c{S}. K.~\"{O}zdemir}, \citenamefont
  {Lee}, \citenamefont {Maier},\ and\ \citenamefont
  {Kim}}]{Kim:QuantumPlasmonics}%
  \BibitemOpen
  \bibfield  {author} {\bibinfo {author} {\bibfnamefont {M.~S.}\ \bibnamefont
  {Tame}}, \bibinfo {author} {\bibfnamefont {K.~R.}\ \bibnamefont {McEnery}},
  \bibinfo {author} {\bibnamefont {\c{S}. K.~\"{O}zdemir}}, \bibinfo {author}
  {\bibfnamefont {J.}~\bibnamefont {Lee}}, \bibinfo {author} {\bibfnamefont
  {S.~A.}\ \bibnamefont {Maier}}, \ and\ \bibinfo {author} {\bibfnamefont
  {M.~S.}\ \bibnamefont {Kim}},\ }\bibfield  {title} {\enquote {\bibinfo
  {title} {Quantum plasmonics},}\ }\href@noop {} {\bibfield  {journal}
  {\bibinfo  {journal} {Nature Phys.}\ }\textbf {\bibinfo {volume} {9}},\
  \bibinfo {pages} {329} (\bibinfo {year} {2013})}\BibitemShut {NoStop}%
\bibitem [{\citenamefont {Hoffmann}\ \emph {et~al.}(2003)\citenamefont
  {Hoffmann}, \citenamefont {Berndt},\ and\ \citenamefont
  {Johansson}}]{Johansson:TwoElectron}%
  \BibitemOpen
  \bibfield  {author} {\bibinfo {author} {\bibfnamefont {G.}~\bibnamefont
  {Hoffmann}}, \bibinfo {author} {\bibfnamefont {R.}~\bibnamefont {Berndt}}, \
  and\ \bibinfo {author} {\bibfnamefont {P.}~\bibnamefont {Johansson}},\
  }\bibfield  {title} {\enquote {\bibinfo {title} {Two-electron photon emission
  from metallic quantum wells},}\ }\href@noop {} {\bibfield  {journal}
  {\bibinfo  {journal} {Phys. Rev. Lett.}\ }\textbf {\bibinfo {volume} {90}},\
  \bibinfo {pages} {046803} (\bibinfo {year} {2003})}\BibitemShut {NoStop}%
\bibitem [{\citenamefont {Tobiska}\ \emph {et~al.}(2006)\citenamefont
  {Tobiska}, \citenamefont {Danon}, \citenamefont {Snyman},\ and\ \citenamefont
  {Nazarov}}]{Nazarov:QuantumTunneling}%
  \BibitemOpen
  \bibfield  {author} {\bibinfo {author} {\bibfnamefont {J.}~\bibnamefont
  {Tobiska}}, \bibinfo {author} {\bibfnamefont {J.}~\bibnamefont {Danon}},
  \bibinfo {author} {\bibfnamefont {I.}~\bibnamefont {Snyman}}, \ and\ \bibinfo
  {author} {\bibfnamefont {Yu.~V.}\ \bibnamefont {Nazarov}},\ }\bibfield
  {title} {\enquote {\bibinfo {title} {Quantum tunneling detection of
  two-photon and two-electron processes},}\ }\href@noop {} {\bibfield
  {journal} {\bibinfo  {journal} {Phys. Rev. Lett.}\ }\textbf {\bibinfo
  {volume} {96}},\ \bibinfo {pages} {096801} (\bibinfo {year}
  {2006})}\BibitemShut {NoStop}%
\bibitem [{\citenamefont {Schneider}\ \emph {et~al.}(2013)\citenamefont
  {Schneider}, \citenamefont {Johansson},\ and\ \citenamefont
  {Berndt}}]{Berndt:HotElectron}%
  \BibitemOpen
  \bibfield  {author} {\bibinfo {author} {\bibfnamefont {N.~L.}\ \bibnamefont
  {Schneider}}, \bibinfo {author} {\bibfnamefont {P.}~\bibnamefont
  {Johansson}}, \ and\ \bibinfo {author} {\bibfnamefont {R.}~\bibnamefont
  {Berndt}},\ }\bibfield  {title} {\enquote {\bibinfo {title} {Hot electron
  cascades in the scanning tunneling microscope},}\ }\href@noop {} {\bibfield
  {journal} {\bibinfo  {journal} {Phys. Rev. B}\ }\textbf {\bibinfo {volume}
  {87}},\ \bibinfo {pages} {045409} (\bibinfo {year} {2013})}\BibitemShut
  {NoStop}%
\bibitem [{\citenamefont {Rammer}\ and\ \citenamefont
  {Smith}(1986)}]{Rammer:QuantumField}%
  \BibitemOpen
  \bibfield  {author} {\bibinfo {author} {\bibfnamefont {J.}~\bibnamefont
  {Rammer}}\ and\ \bibinfo {author} {\bibfnamefont {H.}~\bibnamefont {Smith}},\
  }\bibfield  {title} {\enquote {\bibinfo {title} {Quantum field-theoretical
  methods in transport theory of metals},}\ }\href@noop {} {\bibfield
  {journal} {\bibinfo  {journal} {Rev. Mod. Phys.}\ }\textbf {\bibinfo {volume}
  {58}},\ \bibinfo {pages} {323} (\bibinfo {year} {1986})}\BibitemShut
  {NoStop}%
\bibitem [{\citenamefont {Haug}\ and\ \citenamefont {Jauho}(1998)}]{Jauho}%
  \BibitemOpen
  \bibfield  {author} {\bibinfo {author} {\bibfnamefont {H.}~\bibnamefont
  {Haug}}\ and\ \bibinfo {author} {\bibfnamefont {A.-P.}\ \bibnamefont
  {Jauho}},\ }\href@noop {} {\emph {\bibinfo {title} {Quantum Kinetics in
  Transport and Optics of Semiconductors}}}\ (\bibinfo  {publisher}
  {Springer},\ \bibinfo {address} {Berlin},\ \bibinfo {year}
  {1998})\BibitemShut {NoStop}%
\bibitem [{\citenamefont {Rendell}\ \emph {et~al.}(1978)\citenamefont
  {Rendell}, \citenamefont {Scalapino},\ and\ \citenamefont
  {M{\"u}hlschlegel}}]{Muhlschlegel:Role}%
  \BibitemOpen
  \bibfield  {author} {\bibinfo {author} {\bibfnamefont {R.~W.}\ \bibnamefont
  {Rendell}}, \bibinfo {author} {\bibfnamefont {D.~J.}\ \bibnamefont
  {Scalapino}}, \ and\ \bibinfo {author} {\bibfnamefont {B.}~\bibnamefont
  {M{\"u}hlschlegel}},\ }\bibfield  {title} {\enquote {\bibinfo {title} {Role
  of local plasmon modes in light emission from small-particle tunnel
  junctions},}\ }\href@noop {} {\bibfield  {journal} {\bibinfo  {journal}
  {Phys. Rev. Lett.}\ }\textbf {\bibinfo {volume} {41}},\ \bibinfo {pages}
  {1746} (\bibinfo {year} {1978})}\BibitemShut {NoStop}%
\bibitem [{\citenamefont {Johansson}\ \emph {et~al.}(1990)\citenamefont
  {Johansson}, \citenamefont {Monreal},\ and\ \citenamefont
  {Apell}}]{Apell:Theory}%
  \BibitemOpen
  \bibfield  {author} {\bibinfo {author} {\bibfnamefont {P.}~\bibnamefont
  {Johansson}}, \bibinfo {author} {\bibfnamefont {R.}~\bibnamefont {Monreal}},
  \ and\ \bibinfo {author} {\bibfnamefont {P.}~\bibnamefont {Apell}},\
  }\bibfield  {title} {\enquote {\bibinfo {title} {Theory for light emission
  from a scanning tunneling microscope},}\ }\href@noop {} {\bibfield  {journal}
  {\bibinfo  {journal} {Phys. Rev. B}\ }\textbf {\bibinfo {volume} {42}},\
  \bibinfo {pages} {9210} (\bibinfo {year} {1990})}\BibitemShut {NoStop}%
\bibitem [{\citenamefont {Berndt}\ \emph {et~al.}(1991)\citenamefont {Berndt},
  \citenamefont {Gimzewski},\ and\ \citenamefont
  {Johansson}}]{Johansson:Inelastic}%
  \BibitemOpen
  \bibfield  {author} {\bibinfo {author} {\bibfnamefont {R.}~\bibnamefont
  {Berndt}}, \bibinfo {author} {\bibfnamefont {J.~K.}\ \bibnamefont
  {Gimzewski}}, \ and\ \bibinfo {author} {\bibfnamefont {P.}~\bibnamefont
  {Johansson}},\ }\bibfield  {title} {\enquote {\bibinfo {title} {Inelastic
  tunneling excitation of tip-induced plasmon modes on noble-metal surfaces},}\
  }\href@noop {} {\bibfield  {journal} {\bibinfo  {journal} {Phys. Rev. Lett.}\
  }\textbf {\bibinfo {volume} {67}},\ \bibinfo {pages} {3796} (\bibinfo {year}
  {1991})}\BibitemShut {NoStop}%
\bibitem [{\citenamefont {Garc\'ia-Mart\'in}\ \emph {et~al.}(2011)\citenamefont
  {Garc\'ia-Mart\'in}, \citenamefont {Ward}, \citenamefont {Natelson},\ and\
  \citenamefont {Cuevas}}]{Cuevas:Field}%
  \BibitemOpen
  \bibfield  {author} {\bibinfo {author} {\bibfnamefont {A.}~\bibnamefont
  {Garc\'ia-Mart\'in}}, \bibinfo {author} {\bibfnamefont {D.~R.}\ \bibnamefont
  {Ward}}, \bibinfo {author} {\bibfnamefont {D.}~\bibnamefont {Natelson}}, \
  and\ \bibinfo {author} {\bibfnamefont {J.~C.}\ \bibnamefont {Cuevas}},\
  }\bibfield  {title} {\enquote {\bibinfo {title} {Field enhancement in
  subnanometer metallic gaps},}\ }\href@noop {} {\bibfield  {journal} {\bibinfo
   {journal} {Phys. Rev. B}\ }\textbf {\bibinfo {volume} {83}},\ \bibinfo
  {pages} {193404} (\bibinfo {year} {2011})}\BibitemShut {NoStop}%
\bibitem [{\citenamefont {Johansson}(1998)}]{Johansson:Light}%
  \BibitemOpen
  \bibfield  {author} {\bibinfo {author} {\bibfnamefont {P.}~\bibnamefont
  {Johansson}},\ }\bibfield  {title} {\enquote {\bibinfo {title} {Light
  emission from a scanning tunneling microscope: {Fully} retarded
  calculation},}\ }\href@noop {} {\bibfield  {journal} {\bibinfo  {journal}
  {Phys. Rev. B}\ }\textbf {\bibinfo {volume} {58}},\ \bibinfo {pages} {10823}
  (\bibinfo {year} {1998})}\BibitemShut {NoStop}%
\bibitem [{\citenamefont {Mahan}(2010)}]{Mahan}%
  \BibitemOpen
  \bibfield  {author} {\bibinfo {author} {\bibfnamefont {G.~D.}\ \bibnamefont
  {Mahan}},\ }\href@noop {} {\emph {\bibinfo {title} {Many-particle
  Physics}}},\ \bibinfo {edition} {3rd}\ ed.\ (\bibinfo  {publisher}
  {Springer},\ \bibinfo {year} {2010})\BibitemShut {NoStop}%
\bibitem [{sup()}]{supplemental}%
  \BibitemOpen
  \href@noop {} {}\bibinfo {note} {See Supplemental Material
  [url].}\BibitemShut {Stop}%
\bibitem [{foo({\natexlab{a}})}]{footnote3}%
  \BibitemOpen
  \href@noop {} {}\bibinfo {note} {We use the symbols $S$
  and $\mathcal{S}$ for the correlation function in units of $eV$ and
  $\tfrac{e^2}{s}$, respectively. The replacement $\tfrac{1}{2\pi}
  \leftrightarrow \tfrac{e^2}{h}$ switches between the two.}\BibitemShut
  {Stop}%
\bibitem [{foo({\natexlab{b}})}]{footnote1}%
  \BibitemOpen
  \href@noop {} {}\bibinfo {note} {This follows from the
  relations $S^r - S^a = S^> - S^<$, $S^>(\omega) = S^<(-\omega)$ and
  $D^>(\omega) = D^<(-\omega)$.}\BibitemShut {Stop}%
\bibitem [{\citenamefont {Gavish}\ \emph {et~al.}(2001)\citenamefont {Gavish},
  \citenamefont {Imry},\ and\ \citenamefont {Levinson}}]{Levinson:Quantum}%
  \BibitemOpen
  \bibfield  {author} {\bibinfo {author} {\bibfnamefont {U.}~\bibnamefont
  {Gavish}}, \bibinfo {author} {\bibfnamefont {Y.}~\bibnamefont {Imry}}, \ and\
  \bibinfo {author} {\bibfnamefont {Y.}~\bibnamefont {Levinson}},\ }\bibfield
  {title} {\enquote {\bibinfo {title} {Quantum noise, detailed balance and
  {Kubo} formula in nonequilibrium quantum systems},}\ }in\ \href@noop {}
  {\emph {\bibinfo {booktitle} {Proceedings of the 2001 Recontres de Moriond:
  Electronic Correlations: from Meso- to Nanophysics}}},\ \bibinfo {editor}
  {edited by\ \bibinfo {editor} {\bibfnamefont {T.}~\bibnamefont {Martin}},
  \bibinfo {editor} {\bibfnamefont {G.}~\bibnamefont {Montambaux}}, \ and\
  \bibinfo {editor} {\bibfnamefont {J.}~\bibnamefont {{Tran Thanh Van}}}}\
  (\bibinfo  {publisher} {EDP Science},\ \bibinfo {address} {Lesulis},\
  \bibinfo {year} {2001})\ \bibinfo {note} {arXiv:cond-mat/0211681}\BibitemShut
  {NoStop}%
\bibitem [{\citenamefont {Scalapino}\ \emph {et~al.}(1993)\citenamefont
  {Scalapino}, \citenamefont {White},\ and\ \citenamefont
  {Zhang}}]{Scalapino:TheCriteria}%
  \BibitemOpen
  \bibfield  {author} {\bibinfo {author} {\bibfnamefont {D.~J.}\ \bibnamefont
  {Scalapino}}, \bibinfo {author} {\bibfnamefont {S.~R.}\ \bibnamefont
  {White}}, \ and\ \bibinfo {author} {\bibfnamefont {S.}~\bibnamefont
  {Zhang}},\ }\bibfield  {title} {\enquote {\bibinfo {title} {Insulator, metal,
  or superconductor: {The} criteria},}\ }\href@noop {} {\bibfield  {journal}
  {\bibinfo  {journal} {Phys. Rev. B}\ }\textbf {\bibinfo {volume} {47}},\
  \bibinfo {pages} {7995} (\bibinfo {year} {1993})}\BibitemShut {NoStop}%
\bibitem [{foo({\natexlab{c}})}]{footnote2}%
  \BibitemOpen
  \href@noop {} {}\bibinfo {note} {As opposed to charge
  fluctuations of the contact, $\Delta I = I_\text{tip} + I_\text{sub}$, which
  would be the case if $M_\text{tip}=M_\text{sub}$.}\BibitemShut {Stop}%
\bibitem [{\citenamefont {Blanter}\ and\ \citenamefont
  {B{\"u}ttiker}(2000)}]{Buttiker:ShotNoise}%
  \BibitemOpen
  \bibfield  {author} {\bibinfo {author} {\bibfnamefont {Y.~M.}\ \bibnamefont
  {Blanter}}\ and\ \bibinfo {author} {\bibfnamefont {M.}~\bibnamefont
  {B{\"u}ttiker}},\ }\bibfield  {title} {\enquote {\bibinfo {title} {Shot noise
  in mesoscopic conductors},}\ }\href@noop {} {\bibfield  {journal} {\bibinfo
  {journal} {Phys. Rep.}\ }\textbf {\bibinfo {volume} {336}},\ \bibinfo {pages}
  {1} (\bibinfo {year} {2000})}\BibitemShut {NoStop}%
\bibitem [{foo({\natexlab{d}})}]{footnote4}%
  \BibitemOpen
  \href@noop {} {} \bibinfo {note} {The factor of 4 in
  Eq.~\eqref{eq:Snonint} stems from the fact that the total noise, $S^< =
  \sum_{\alpha\beta} S_{\alpha\beta}^<$, is considered.}\BibitemShut {Stop}%
\bibitem [{\citenamefont {Zee}(2010)}]{Zee}%
  \BibitemOpen
  \bibfield  {author} {\bibinfo {author} {\bibfnamefont {A.}~\bibnamefont
  {Zee}},\ }\href@noop {} {\emph {\bibinfo {title} {Quantum Field Theory in a
  Nutshell}}},\ \bibinfo {edition} {2nd}\ ed.\ (\bibinfo  {publisher}
  {Princeton University Press},\ \bibinfo {address} {Princeton},\ \bibinfo
  {year} {2010})\BibitemShut {NoStop}%
\bibitem [{\citenamefont {Scholl}\ \emph {et~al.}(2013)\citenamefont {Scholl},
  \citenamefont {Garc\'ia-Etxarri}, \citenamefont {Koh},\ and\ \citenamefont
  {Dionne}}]{Dionne:Observation}%
  \BibitemOpen
  \bibfield  {author} {\bibinfo {author} {\bibfnamefont {J.~A.}\ \bibnamefont
  {Scholl}}, \bibinfo {author} {\bibfnamefont {A.}~\bibnamefont
  {Garc\'ia-Etxarri}}, \bibinfo {author} {\bibfnamefont {A.~Leen}\ \bibnamefont
  {Koh}}, \ and\ \bibinfo {author} {\bibfnamefont {J.~A.}\ \bibnamefont
  {Dionne}},\ }\bibfield  {title} {\enquote {\bibinfo {title} {Observation of
  quantum tunneling between two plasmonic nanoparticles},}\ }\href@noop {}
  {\bibfield  {journal} {\bibinfo  {journal} {Nano. Lett.}\ }\textbf {\bibinfo
  {volume} {13}},\ \bibinfo {pages} {564} (\bibinfo {year} {2013})}\BibitemShut
  {NoStop}%
\bibitem [{\citenamefont {Song}\ \emph {et~al.}(2011)\citenamefont {Song},
  \citenamefont {Nordlander},\ and\ \citenamefont {Gao}}]{Gao:QMStudy}%
  \BibitemOpen
  \bibfield  {author} {\bibinfo {author} {\bibfnamefont {P.}~\bibnamefont
  {Song}}, \bibinfo {author} {\bibfnamefont {P.}~\bibnamefont {Nordlander}}, \
  and\ \bibinfo {author} {\bibfnamefont {S.}~\bibnamefont {Gao}},\ }\bibfield
  {title} {\enquote {\bibinfo {title} {Quantum mechanical study of the coupling
  of plasmon excitations to atomic-scale electron transport},}\ }\href@noop {}
  {\bibfield  {journal} {\bibinfo  {journal} {J. Chem. Phys.}\ }\textbf
  {\bibinfo {volume} {134}},\ \bibinfo {pages} {074701} (\bibinfo {year}
  {2011})}\BibitemShut {NoStop}%
\bibitem [{\citenamefont {Esteban}\ \emph {et~al.}(2012)\citenamefont
  {Esteban}, \citenamefont {Borisov}, \citenamefont {Nordlander},\ and\
  \citenamefont {Aizpurua}}]{Aizpurua:Bridging}%
  \BibitemOpen
  \bibfield  {author} {\bibinfo {author} {\bibfnamefont {R.}~\bibnamefont
  {Esteban}}, \bibinfo {author} {\bibfnamefont {A.~G.}\ \bibnamefont
  {Borisov}}, \bibinfo {author} {\bibfnamefont {P.}~\bibnamefont {Nordlander}},
  \ and\ \bibinfo {author} {\bibfnamefont {Javier}\ \bibnamefont {Aizpurua}},\
  }\bibfield  {title} {\enquote {\bibinfo {title} {Bridging quantum and
  classical plasmonics with a quantum-corrected model},}\ }\href@noop {}
  {\bibfield  {journal} {\bibinfo  {journal} {Nature Comm.}\ }\textbf {\bibinfo
  {volume} {3}},\ \bibinfo {pages} {825} (\bibinfo {year} {2012})}\BibitemShut
  {NoStop}%
\bibitem [{\citenamefont {Mortensen}\ \emph {et~al.}(2014)\citenamefont
  {Mortensen}, \citenamefont {Raza}, \citenamefont {Wubs}, \citenamefont
  {S{\o}ndergaard},\ and\ \citenamefont
  {Bozhevolnyi}}]{Bozhevolnyi:Generalized}%
  \BibitemOpen
  \bibfield  {author} {\bibinfo {author} {\bibfnamefont {N.~Asger}\
  \bibnamefont {Mortensen}}, \bibinfo {author} {\bibfnamefont {S.}~\bibnamefont
  {Raza}}, \bibinfo {author} {\bibfnamefont {M.}~\bibnamefont {Wubs}}, \bibinfo
  {author} {\bibfnamefont {T.}~\bibnamefont {S{\o}ndergaard}}, \ and\ \bibinfo
  {author} {\bibfnamefont {S.~I.}\ \bibnamefont {Bozhevolnyi}},\ }\bibfield
  {title} {\enquote {\bibinfo {title} {A generalized non-local optical response
  theory for plasmonic nanostructures},}\ }\href@noop {} {\bibfield  {journal}
  {\bibinfo  {journal} {Nature Comm.}\ }\textbf {\bibinfo {volume} {5}},\
  \bibinfo {pages} {3809} (\bibinfo {year} {2014})}\BibitemShut {NoStop}%
\bibitem [{\citenamefont {Teperik}\ \emph
  {et~al.}(2013{\natexlab{a}})\citenamefont {Teperik}, \citenamefont
  {Nordlander}, \citenamefont {Aizpurua},\ and\ \citenamefont
  {Borisov}}]{Borisov:Robust}%
  \BibitemOpen
  \bibfield  {author} {\bibinfo {author} {\bibfnamefont {T.~V.}\ \bibnamefont
  {Teperik}}, \bibinfo {author} {\bibfnamefont {P.}~\bibnamefont {Nordlander}},
  \bibinfo {author} {\bibfnamefont {J.}~\bibnamefont {Aizpurua}}, \ and\
  \bibinfo {author} {\bibfnamefont {A.~G.}\ \bibnamefont {Borisov}},\
  }\bibfield  {title} {\enquote {\bibinfo {title} {Robust subnanometric plasmon
  ruler by rescaling of the nonlocal optical response},}\ }\href@noop {}
  {\bibfield  {journal} {\bibinfo  {journal} {Phys. Rev. Lett.}\ }\textbf
  {\bibinfo {volume} {110}},\ \bibinfo {pages} {263901} (\bibinfo {year}
  {2013}{\natexlab{a}})}\BibitemShut {NoStop}%
\bibitem [{\citenamefont {Teperik}\ \emph
  {et~al.}(2013{\natexlab{b}})\citenamefont {Teperik}, \citenamefont
  {Nordlander}, \citenamefont {Aizpurua},\ and\ \citenamefont
  {Borisov}}]{Borisov:Quantum}%
  \BibitemOpen
  \bibfield  {author} {\bibinfo {author} {\bibfnamefont {T.~V.}\ \bibnamefont
  {Teperik}}, \bibinfo {author} {\bibfnamefont {P.}~\bibnamefont {Nordlander}},
  \bibinfo {author} {\bibfnamefont {J.}~\bibnamefont {Aizpurua}}, \ and\
  \bibinfo {author} {\bibfnamefont {A.~G.}\ \bibnamefont {Borisov}},\
  }\bibfield  {title} {\enquote {\bibinfo {title} {Quantum effects and
  nonlocality in strongly coupled plasmonic nanowire dimers},}\ }\href@noop {}
  {\bibfield  {journal} {\bibinfo  {journal} {Optics Express}\ }\textbf
  {\bibinfo {volume} {21}},\ \bibinfo {pages} {27306} (\bibinfo {year}
  {2013}{\natexlab{b}})}\BibitemShut {NoStop}%
\bibitem [{\citenamefont {Aizpurua}\ \emph {et~al.}(2002)\citenamefont
  {Aizpurua}, \citenamefont {Hoffmann}, \citenamefont {Apell},\ and\
  \citenamefont {Berndt}}]{Berndt:Electromagnetic}%
  \BibitemOpen
  \bibfield  {author} {\bibinfo {author} {\bibfnamefont {J.}~\bibnamefont
  {Aizpurua}}, \bibinfo {author} {\bibfnamefont {G.}~\bibnamefont {Hoffmann}},
  \bibinfo {author} {\bibfnamefont {S.~P.}\ \bibnamefont {Apell}}, \ and\
  \bibinfo {author} {\bibfnamefont {R.}~\bibnamefont {Berndt}},\ }\bibfield
  {title} {\enquote {\bibinfo {title} {Electromagnetic coupling on an atomic
  scale},}\ }\href@noop {} {\bibfield  {journal} {\bibinfo  {journal} {Phys.
  Rev. Lett.}\ }\textbf {\bibinfo {volume} {89}},\ \bibinfo {pages} {156803}
  (\bibinfo {year} {2002})}\BibitemShut {NoStop}%
\end{thebibliography}%

\pagebreak
\widetext
\clearpage


\section*{Supplementary material}

# Supplemental material for "Theory of light emission from quantum noise in plasmonic contacts: above-threshold emission from higher-order electron-plasmon scattering"

Kristen Kaasbjerg[1,2] and Abraham Nitzan[2]

[1]*Department of Condensed Matter Physics, Weizmann Institute of Science, Rehovot, Israel 76100*
[2]*School of Chemistry, The Sackler Faculty of Exact Sciences, Tel Aviv University, Tel Aviv 69978, Israel*
(Dated: January 25, 2015)

## I. LSP GREEN'S FUNCTION

The central object of interest for our description of plasmonic light emission from biased STM contacts, is the contour-ordered GF for the localized surface-plasmon polariton (LSP) of the contact represented by the quantized vector potential

$$\mathbf{A}(\mathbf{r}) = \boldsymbol{\xi}_{\text{pl}}(\mathbf{r})\sqrt{\frac{\hbar}{2\Omega\epsilon_0\omega_{\text{pl}}}}\left(a^\dagger + a\right), \tag{1}$$

where $\Omega$ is a quantization volume and $\boldsymbol{\xi}_{\text{pl}}$ is the mode vector. The contour-ordered GF is defined by

$$D(\tau,\tau') = -i\langle T_c A(\tau)A(\tau')\rangle, \tag{2}$$

where $A = a + a^\dagger$ and $T_c$ is the time-ordering operator on the Keldysh contour. In the presence of interactions, it obeys the Dyson equation

$$D(\tau,\tau') = d_0(\tau,\tau') + \int d\tau_1 \int d\tau_2 \\ \times d_0(\tau,\tau_1)\Pi(\tau_1,\tau_2)D(\tau_2,\tau'), \tag{3}$$

where $d_0$ is the *bare* GF and $\Pi$ is the *irreducible* self-energy.

### A. Electron-plasmon self-energy

In order to establish an *exact* expression for the el-pl self-energy, we start from the perturbation expansion of the LSP GF in terms of the $S$-matrix on the Keldysh contour,

$$S_c = T_c \exp\left[-i\int_c d\tau_1 V(\tau_1)\right], \tag{4}$$

where

$$V = \int d\mathbf{r}\,\mathbf{j}_{\text{el}}(\mathbf{r})\cdot\mathbf{A}(\mathbf{r}) \tag{5}$$

is the el-pl interaction accounting for the interaction with the tunnel current, $\mathbf{j}_{\text{el}} = \mathbf{j}^\nabla + \mathbf{j}^A$, which is a sum of paramagnetic and diamagnetic components. As explained in the main text, we here neglect the diamagnetic component and focus on the paramagnetic self-energy, $\Pi^\nabla$, which governs the excitation dynamics of the LSP.

Introducing the current operator

$$j = \sqrt{\frac{\hbar}{2V\epsilon_0\omega_{\text{pl}}}} \int d\mathbf{r}\,\boldsymbol{\xi}_{\text{pl}}(\mathbf{r})\cdot\mathbf{j}^\nabla(\mathbf{r}), \tag{6}$$

the $S$-matrix expansion of the LSP GF in the paramagnetic interaction can be written

$$\begin{aligned}
D(\tau,\tau') &= -i\langle T_c S_c A(\tau)A(\tau')\rangle_0 \\
&= -i\sum_{n=0}^\infty \frac{(-i)^n}{n!} \int d\tau_1 \cdots \int d\tau_n \langle T_c A(\tau)V(\tau_1)\cdots V(\tau_n)A(\tau')\rangle_0^{\text{con}} \\
&= -i\sum_{n=0}^\infty \frac{(-i)^{2n}}{(2n)!} \int d\tau_1 \cdots \int d\tau_{2n} \langle T_c A(\tau)A(\tau_1)\cdots A(\tau_{2n})A(\tau')\rangle_0 \langle T_c j(\tau_1)\cdots j(\tau_{2n})\rangle_0,
\end{aligned} \tag{7}$$

where the sum is over connected diagrams and, in the last equality, we have replaced $n \to 2n$ (the expectation value of an odd number of boson operators is zero, implying that only even orders contribute in the perturbation series).

In order to identify the bosonic self-energy from the perturbation series, the expectation value of the time-ordered

bosonic operators is rewritten as

$$\langle T_c A(\tau)A(\tau_1)\cdots A(\tau_{2n})A(\tau')\rangle_0 = 2n(2n-1)\langle T_c A(\tau)A(\tau_1)\rangle_0 \langle T_c A(\tau_2)\cdots A(\tau_{2n-1})\rangle_0 \langle T_c A(\tau_{2n})A(\tau')\rangle_0 \qquad (8)$$

The factor of $2n(2n-1)$ on the right-hand side comes from number of ways two internal times can be paired up with the two external times $\tau, \tau'$ in the Wick's contraction of the bosonic expectation value. For each pairing $\tau, \tau_n$ and $\tau', \tau_{n'}$, the contractions of the remaining $2n-2$ internal boson operators are recollected in the uncontracted expectation value.

The GF can now be written

$$D(\tau, \tau') = d_0(\tau, \tau') + \int d\tau_1 \int d\tau_2\, d_0(\tau,\tau_1)\Pi^\nabla(\tau_1,\tau_2)d_0(\tau_2,\tau') \qquad (9)$$

where the *reducible* self-energy is defined by

$$\begin{aligned}\Pi^\nabla(\tau,\tau') &= i\sum_{n=1}^\infty \frac{(-i)^{2n}}{(2n)!} 2n(2n-1) \int d\tau_1 \cdots \int d\tau_{2n-2}\, \langle T_c j(\tau)j(\tau_1)\cdots j(\tau_{2n-2})j(\tau')\rangle_0 \langle T_c A(\tau_1)\cdots A(\tau_{2n-2})\rangle_0 \\ &= -i\sum_{n=0}^\infty \frac{(-i)^{2n}}{(2n)!} \int d\tau_1 \cdots \int d\tau_{2n}\, \langle T_c j(\tau)j(\tau_1)\cdots j(\tau_{2n})j(\tau')\rangle_0 \langle T_c A(\tau_1)\cdots A(\tau_{2n})\rangle_0 \\ &\equiv -i S(\tau,\tau').\end{aligned} \qquad (10)$$

Here, we have identified the sum over connected diagrams in the second line as the perturbation expansion of the correlation function $S(\tau,\tau') = \langle T_c \delta j(\tau)\delta j(\tau')\rangle$ where $\delta j = j - \langle j \rangle$. Writing out the correlation function we have $S(\tau,\tau') = \langle T_c j(\tau) j(\tau')\rangle - \langle j \rangle^2$. Here, the last term cancels the disconnected diagrams from the perturbation expansion of the first term which do not appear in the expansion for the LSP GF in Eq. (7).

From the above, it follows trivially that the *irreducible* self-energy is given by

$$\Pi^\nabla(\tau,\tau') \equiv -iS^{\text{irr}}(\tau,\tau'), \qquad (11)$$

where $S^{\text{irr}}$ is the *irreducible* part of the correlation function $S$.

### B. Self-energy for the contact model

For the simple contact model considered in the main text, the paramagnetic part of the interaction can be written on the form[1]

$$V = \sum_\alpha M_\alpha I_\alpha(a^\dagger + a) \qquad (12)$$

where $M_\alpha$ is the coupling constant at lead $\alpha$ and the paramagnetic current operator is defined by

$$I_\alpha = i\sum_k \left[t_\alpha c_{\alpha k}^\dagger d - \text{h.c}\right], \qquad (13)$$

where the sum is over states $k$ in the lead. It is here assumed that the tunnel coupling $t_\alpha$ and the coupling constant $M_\alpha$ are independent on the state index $k$.

Repeating the perturbation expansion for the plasmonic GF in the preceding section with the interaction in Eq. (12), we find that the paramagnetic self-energy can be written as a sum over lead components, $\Pi^\nabla = \sum_{\alpha\beta} \Pi^\nabla_{\alpha\beta}$, where

$$\Pi^\nabla_{\alpha\beta}(\tau,\tau') = -iM_\alpha M_\beta S^{\text{irr}}_{\alpha\beta}(\tau,\tau'), \qquad (14)$$

and $S_{\alpha\beta}(\tau,\tau') = \langle T_c \delta I_\alpha(\tau) \delta I_\beta(\tau')\rangle$ is the *reducible* correlation function with $\delta I_\alpha = I_\alpha - \langle I_\alpha\rangle$.

#### 1. Identities for the self-energy

From the hermitian property of the paramagnetic current operator $I_\alpha$, the following set of identities between the different lead-lead components of the retarded/advanced and lesser/greater self-energies can be derived.

For the retarded/advanced components we have

$$\Pi^r_{\alpha\beta}(\omega) = \left[\Pi^a_{\beta\alpha}(\omega)\right]^* = \Pi^a_{\beta\alpha}(-\omega) = \left[\Pi^r_{\alpha\beta}(-\omega)\right]^*. \qquad (15)$$

For the lesser/greater components,

$$\Pi^<_{\alpha\beta}(\omega) = \Pi^>_{\beta\alpha}(-\omega). \qquad (16)$$

In addition, the components of the lesser self-energy in frequency domain are related as

$$[\Pi^<_{\alpha\beta}(\omega)]^* = -\Pi^<_{\beta\alpha}(\omega). \qquad (17)$$

Hence, the diagonals $\Pi^<_{\alpha\alpha}$ are purely imaginary while the off-diagonal elements, in general, have both a real and an imaginary part. Note, however, that the real parts of the off-diagonal components cancel each other, implying that the total lesser self-energy becomes purely imaginary.

## II. PERTURBATION SERIES FOR THE SELF-ENERGY

The perturbation series and the rules for evaluating the corresponding Feynman diagrams are most easily developed by writing the el-pl interaction (12) on the general form

$$V = \sum_{ij} M_{ij} c_i^\dagger c_j (a^\dagger + a) \qquad (18)$$

where $i, j = \alpha, d$ and $\alpha = L, R$ is a composite lead/state index, $\alpha = (\alpha, k)$. With the interaction written on this form, the el-pl interaction (12) can be represented by the coupling matrix

$$\mathbf{M} = i \begin{pmatrix} 0 & -t_L M_L & -t_R M_R \\ t_L M_L & 0 & 0 \\ t_R M_R & 0 & 0 \end{pmatrix}, \qquad (19)$$

in the $(d, L, R)$ basis.

We can now write up the perturbation series for the lead-lead components of the self-energy and get

$$\Pi_{\alpha\beta}^\nabla(\tau,\tau') = -i M_\alpha M_\beta \sum_{n=0}^\infty \frac{(-i)^n}{n!} \int d\tau_1 \cdots \int d\tau_n \langle T I_\alpha(\tau) V(\tau_1) \cdots V(\tau_n) I_\beta(\tau') \rangle_0^{\text{con}}$$

$$= -i \sum_{ij \in \{\alpha,d\}} \sum_{i'j' \in \{\beta,d\}} M_{ij} M_{i'j'} \sum_{n=0}^\infty \frac{(-i)^{2n}}{(2n)!} \int d\tau_1 \cdots \int d\tau_{2n} \sum_{i_1 j_1 \cdots i_{2n} j_{2n}} M_{i_1 j_1} \cdots M_{i_{2n} j_{2n}}$$

$$\times \langle T_c c_i^\dagger(\tau) c_j(\tau) c_{i_1}^\dagger(\tau_1) c_{j_1}(\tau_1) \cdots c_{i_{2n}}^\dagger(\tau_{2n}) c_{j_{2n}}(\tau_{2n}) c_{i'}^\dagger(\tau') c_{j'}(\tau') \rangle_0 \langle T_c A(\tau_1) \cdots A(\tau_{2n}) \rangle_0. \qquad (20)$$

Here the first two sums over $i,j$ and $i',j'$ are associated with the current operators appearing explicitly in the correlation function $S_{\alpha\beta}$.

### A. The contact GF

Given the structure of the perturbation series in (20) above, the fundamental building block in the Feynman diagrams for the self-energy is identified as the contact Green's function defined by

$$G_{ij}(\tau, \tau') = -i \langle T_c c_i(\tau) c_j^\dagger(\tau') \rangle_0, \qquad (21)$$

where $i, j = \alpha, d$.

In the absence of interactions, i.e. only tunneling between the leads and the level is included, the components of the contact GF are given by

$$G_{dd}(\tau, \tau') = g_d(\tau, \tau')$$
$$+ \int d\tau_1 \int d\tau_2 \, g_d(\tau, \tau_1) \Sigma(\tau_1, \tau_2) G_{dd}(\tau_2, \tau') \qquad (22)$$

$$G_{\alpha d}(\tau, \tau') = \int d\tau_1 \, t_\alpha g_\alpha(\tau, \tau_1) G_{dd}(\tau_1, \tau') \qquad (23)$$

$$G_{d\alpha}(\tau, \tau') = \int d\tau_1 \, G_{dd}(\tau, \tau_1) t_\alpha^* g_\alpha(\tau_1, \tau') \qquad (24)$$

$$G_{\alpha\beta}(\tau, \tau') = \delta_{\alpha\beta} g_\alpha(\tau, \tau')$$
$$+ \int d\tau_1 \int d\tau_2 \, g_\alpha(\tau, \tau_1) t_\alpha G_{dd}(\tau_1, \tau_2) t_\alpha^* g_\beta(\tau_2, \tau') \qquad (25)$$

where $\Sigma = \Sigma_L + \Sigma_R$, $\Sigma_\alpha = \sum_k |t_\alpha|^2 g_\alpha$ is the self-energy due to the coupling to the leads here described within the wide-band limit, and $g_{\alpha/d}$ is the *bare* lead/dot GFs.

### B. Feynman rules

The perturbation series for the self-energy in (20) can be represented by the Feynman diagrams familiar from perturbation expansions of other two-particle GFs in the presence of an electron-boson interaction (see Fig. 2 of the main text)[2]. Here, we give the Feynman rules that apply to the present case:

- At order $2n$, draw $2n+2$ vertices.

- Label each vertex with $i, j$ indices and multiply by $M_{ij}$.

- Connect vertices with the components of the contact GFs (21) matching the in/out going vertex indices.

- Connect internal vertices with *bare* boson GFs.

- At each vertex, sum over the states of the involved lead. The appearence of $\sum_k t_\alpha / \sum_k t_\alpha^*$ at each vertex, implies that the component of the contact GF entering/exiting the vertex with a lead index can be replaced according to:

  $$- G_{\alpha\beta}(\tau, \tau') \to \delta_{\alpha\beta} \Sigma_\alpha(\tau, \tau')$$
  $$+ \int d\tau_1 \int d\tau_2 \, \Sigma_\alpha(\tau, \tau_1) G_{dd}(\tau_1, \tau_2) \Sigma_\beta(\tau_2, \tau')$$

- $G_{\alpha d}(\tau, \tau') \to \int d\tau_1 \Sigma_\alpha(\tau, \tau_1) G_{dd}(\tau_1, \tau')$
- $G_{d\alpha}(\tau, \tau') \to \int d\tau_1 G_{dd}(\tau, \tau_1) \Sigma_\alpha(\tau_1, \tau')$,

i.e., only the $k$-summed lead-dot/dot-lead/lead-lead GFs appear in the perturbation series.

- Integrate over all internal contour times.
- Factor of $-1$ for each fermion loop.
- Factor of $-i \times i^{n+2}$ at order $2n$.

The evaluation of the self-energy (retarded/advanced and lesser/greater) on the real-time axis can be accomplished using Keldysh GFs[3].

### III. GFS IN KELDYSH SPACE

In a perturbative calculation of the self-energy, it can be advantageous to work with the GFs in Keldysh space spanned by the forward ($-$) and backward ($+$) branches of the Keldysh contour[3]. The GFs are here expressed as $2 \times 2$ matrices (indicated by the $\check{\ }$ sign in the following),

$$\check{G} = \begin{pmatrix} G^{--} & G^{-+} \\ G^{+-} & G^{++} \end{pmatrix}, \quad (26)$$

with the components corresponding to the different combinations for the positions of the two time arguments on the contour.

This formulation formally arises from the contour integrals which can be recast to real-time integrals as $\int_c d\tau_i \to \int_- dt_i^- + \int_+ dt_i^+ = \int dt_i^- - \int dt_i^+$. For an expression on the contour like

$$X(\tau, \tau') = \int_c d\tau_1 d\tau_2 \cdots d\tau_n \\ \times A_1(\tau, \tau_1) A_2(\tau_1, \tau_2) \cdots A_n(\tau_n, \tau') \quad (27)$$

which contain $n$ contour integrals, the different real-time components in Keldysh space can hence be obtained as

$$X^{\sigma\sigma'}(t, t') = \sum_{\sigma_1 \sigma_2 \ldots \sigma_n} \eta_1 \eta_2 \cdots \eta_n \int dt_1 dt_2 \cdots dt_n \\ \times A_1^{\sigma\sigma_1}(t, t_1) A_2^{\sigma_1\sigma_2}(t_1, t_2) \cdots A_n^{\sigma_n\sigma'}(t_n, t'), \quad (28)$$

where $\sigma_i = -/+$ is the contour branch index and $\eta_i = +/-$ is the accompanying sign. The prefactor in front of the integrals hence keeps track of the overall sign arising from the contour variables residing on the backward branch.

The GF in Keldysh space obeys the Dyson equation

$$\check{G} = \check{g}_0 + \check{g}_0 \check{\sigma}_3 \check{\Sigma} \check{\sigma}_3 \check{G}, \quad (29)$$

where $\check{\sigma}_i$ denotes the Pauli matrices, and can hence be obtained in the usual way as

$$\check{G}^{-1} = \check{g}_0^{-1} - \check{\sigma}_3 \check{\Sigma} \check{\sigma}_3, \quad (30)$$

where $g_0$ is the *bare* GF and $\Sigma$ is the self-energy accounting for interactions and couplings to, e.g., external leads. The self-energy is defined as

$$\check{\Sigma} = \begin{pmatrix} \Sigma^{--} & \Sigma^{-+} \\ \Sigma^{+-} & \Sigma^{++} \end{pmatrix}, \quad (31)$$

and the matrix product with the Pauli matrices in (30) adds a minus sign to the off-diagonal elements in order to account for the above-mentioned sign arising from the contour integration in the Dyson equation.

When the *bare* GF $\check{g}_0$ is diagonal, the full GF is given by

$$\check{G} = \frac{1}{\mathcal{G}} \times \begin{pmatrix} -(g_0^{++})^{-1} + \Sigma^{++} & \Sigma^{-+} \\ \Sigma^{+-} & -(g_0^{--})^{-1} + \Sigma^{--} \end{pmatrix}, \quad (32)$$

where the denominator of the prefactor is given by

$$\mathcal{G} = \Sigma^{-+}\Sigma^{+-} - \left[(g_0^{--})^{-1} - \Sigma^{--}\right]\left[(g_0^{++})^{-1} - \Sigma^{++}\right] \quad (33)$$

For cases where $\Sigma^{--} = \Sigma^{++}$ and $g_0^{--} = -g_0^{++}$ (see, e.g., below), the denominator of the prefactor simplifies to $\mathcal{G} = \Sigma^r \Sigma^a - (g_0^{--})^{-1}(g_0^{++})^{-1}$.

The above also holds for a bosonic GFs.

In the subsections below, we give expressions for some of the Keldysh space GFs relevant for the present work.

#### A. GF for electronic level coupled to leads

For a single electronic level the, the GF in the absence of coupling to leads and interactions is given by[2]

$$\check{g}_0(\varepsilon) = \begin{pmatrix} g^{--}(\varepsilon) & 0 \\ 0 & g^{++}(\varepsilon) \end{pmatrix}. \quad (34)$$

where $g^{--}(\varepsilon) = \frac{1}{\varepsilon - \varepsilon_0 + i\,\mathrm{sgn}(\varepsilon)}$ and $g^{++}(\varepsilon) = -g^{--}(\varepsilon)$.

The self-energy due to coupling to leads is

$$\check{\Sigma}_\alpha(\varepsilon) = \begin{pmatrix} \Lambda_\alpha(\varepsilon) & 0 \\ 0 & -\Lambda_\alpha(\varepsilon) \end{pmatrix} + i \begin{pmatrix} \Gamma_\alpha(\varepsilon)[f_\alpha(\varepsilon) - 1/2] & -\Gamma_\alpha(\varepsilon) f_\alpha(\varepsilon) \\ \Gamma_\alpha(\varepsilon)[1 - f_\alpha(\varepsilon)] & \Gamma_\alpha(\varepsilon)[f_\alpha(\varepsilon) - 1/2] \end{pmatrix}. \quad (35)$$

With this, we can calculate the GF $G_0$ for the coupled level and find

$$\check{G}_0(\varepsilon) = \frac{1}{\mathcal{G}(\varepsilon)} \times \begin{pmatrix} \varepsilon - \varepsilon_0 - \sum_\alpha \Lambda_\alpha(\varepsilon) + i\sum_\alpha \Gamma_\alpha(\varepsilon)\left[f_\alpha(\varepsilon) - 1/2\right] & i\sum_\alpha \Gamma_\alpha(\varepsilon) f_\alpha(\varepsilon) \\ -i\sum_\alpha \Gamma_\alpha(\varepsilon)\left[1 - f_\alpha(\varepsilon)\right] & -\varepsilon + \varepsilon_0 + \sum_\alpha \Lambda_\alpha(\varepsilon) + i\sum_\alpha \Gamma_\alpha(\varepsilon)\left[f_\alpha(\varepsilon) - 1/2\right] \end{pmatrix}. \quad (36)$$

Note that the off diagonals are given by the usual lesser and greater GFs as expected. In the wide-band limit the denominator of the prefactor reduces to $\mathcal{G}(\varepsilon) = (\varepsilon - \varepsilon_0)^2 + (\Gamma/2)^2$.

### B. GF for noninteracting lead

The Keldysh GF for noninteracting lead fermions in equilibrium is given by[2]

$$\check{g}_k(\varepsilon) = P\frac{1}{\varepsilon - \varepsilon_k}\begin{pmatrix} 1 & 0 \\ 0 & -1 \end{pmatrix} + 2\pi i\delta(\varepsilon - \varepsilon_k)\begin{pmatrix} f(\varepsilon) - 1/2 & f(\varepsilon) \\ -(1-f(\varepsilon)) & f(\varepsilon) - 1/2 \end{pmatrix}. \quad (37)$$

In the wideband limit, the real part is thrown away.

### C. Level-lead GF

The contour-ordered level-lead GF is given by

$$G_{\alpha d}(\tau, \tau') = \int d\tau_1\, g_\alpha(\tau, \tau_1) t_\alpha G_0(\tau_1, \tau'). \quad (38)$$

In Keldysh space it can be written on the form

$$\check{G}_{\alpha d}(\varepsilon) = t_\alpha \check{g}_\alpha(\varepsilon) \check{\sigma}_3 \check{G}_0(\varepsilon). \quad (39)$$

When multiplied by $t_\alpha^*$ and summed over $k$ this can be expressed in terms of the coupling self-energy as

$$\sum_k t_\alpha^* \check{G}_{\alpha d}(\varepsilon) = \check{\Sigma}_\alpha(\varepsilon)\check{\sigma}_3 \check{G}_0(\varepsilon). \quad (40)$$

### D. GF for damped plasmon

The *bare* GF for a plasmon with frequency $\omega_0$ is given by

$$\check{d}_0(\omega) = \begin{pmatrix} d_0^{--}(\omega) & 0 \\ 0 & d_0^{++}(\omega) \end{pmatrix} \quad (41)$$

where $d_0^{--}(\omega) = \frac{1}{\omega - \omega_0 + i\delta} - \frac{1}{\omega + \omega_0 - i\delta} = \frac{2\omega_0}{\omega^2 - \omega_0^2 + i\delta}$ and $d_0^{++}(\omega) = -d_0^{--}(\omega)$.

The self-energy due to damping mechanisms described by a phenomenological damping rate $\gamma$ is given by

$$\check{\Pi}_{\text{damp}}(\omega) = \begin{pmatrix} -i\gamma\left[F(\omega) + \text{sgn}(\omega)/2\right] & i\gamma F(\omega) \\ i\gamma F(-\omega) & -i\gamma\left[F(\omega) + \text{sgn}(\omega)/2\right] \end{pmatrix} \quad (42)$$

where $F(\omega) = |n_B(\omega)|$ and $n_B(x) = (e^{\beta x} - 1)^{-1}$ is the Bose-Einstein distribution function ($\beta = 1/k_B T$). For the damped plasmon GF, we then find

$$\check{D}_0(\omega) = \frac{2\omega_0}{\mathcal{D}(\omega)}\begin{pmatrix} \omega^2 - \omega_0^2 - 2\omega_0 i\gamma\left[F(\omega) + \text{sgn}(\omega)/2\right] & -2\omega_0 i\gamma F(\omega) \\ -2\omega_0 i\gamma F(-\omega) & -\omega^2 + \omega_0^2 - 2\omega i\gamma\left[F(\omega) + \text{sgn}(\omega)/2\right] \end{pmatrix}, \quad (43)$$

where $\mathcal{D}(\omega) = (\omega^2 - \omega_0^2)^2 + 4\omega_0^2(\gamma/2)^2$.

## IV. EXPRESSION FOR THE NONINTERACTING QUANTUM NOISE

In the following, we derive an expression for the noninteracting finite-frequency noise for the single-channel contact considered in the main text. We consider here the total noise

$$S^< = S_{LL}^< + S_{RR}^< - S_{LR}^< - S_{RL}^<, \quad (44)$$

and note that the current-current correlation function $S^<(t, t') = \langle I(t')I(t)\rangle$ is the lesser component of the contour-ordered correlation function $S(\tau, \tau') = \langle T_c I(\tau) I(\tau')\rangle$.

The *noninteracting* quantum noise is given by the

lesser component of the *bare* bubble diagram. Using the Feynman rules for the self-energy above, we can write up the expression for its lead-lead components and obtain

$$\begin{aligned}S_{\alpha\beta}(\tau,\tau') &= \langle T_c I_\alpha(\tau) I_\beta(\tau')\rangle_0 \\ &= -\left[t_\alpha^* t_\beta^* G_{\alpha d}(\tau,\tau') G_{\beta d}(\tau',\tau) + t_\alpha t_\beta G_{d\beta}(\tau,\tau') G_{d\alpha}(\tau',\tau) - t_\alpha^* t_\beta G_{\alpha\beta}(\tau,\tau') G_{dd}(\tau',\tau) - t_\alpha t_\beta^* G_{dd}(\tau,\tau') G_{\beta\alpha}(\tau',\tau)\right] \\ &= \left(\delta_{\alpha\beta}\Sigma_\alpha(\tau,\tau') + \int d\tau_1 \int d\tau_2\, \Sigma_\alpha(\tau,\tau_1) G_{dd}(\tau_1,\tau_2)\Sigma_\beta(\tau_2,\tau')\right) G_{dd}(\tau',\tau) \\ &\quad + G_{dd}(\tau,\tau')\left(\delta_{\alpha\beta}\Sigma_\alpha(\tau',\tau) + \int d\tau_1 \int d\tau_2\, \Sigma_\beta(\tau',\tau_1) G_{dd}(\tau_1,\tau_2)\Sigma_\alpha(\tau_2,t)\right) \\ &\quad - \int d\tau_1\, \Sigma_\alpha(\tau,\tau_1) G_{dd}(\tau_1,\tau') \int d\tau_1\, \Sigma_\beta(\tau',\tau_1) G_{dd}(\tau_1,t) - \int d\tau_1\, G_{dd}(\tau,\tau_1)\Sigma_\beta(\tau_1,\tau') \int d\tau_1\, G_{dd}(\tau',\tau_1)\Sigma_\alpha(\tau_1,t).\end{aligned}$$
(45)

Taking the lesser component of this expression, carrying out the transition to the real-time axis (using either Langreth conversion rules[4] or Keldysh GFs) and moving to frequency domain, we find that the finite-frequency noise can be expressed in terms of the energy-dependent transmission function

$$T(\varepsilon) = \frac{\Gamma_L \Gamma_R}{(\varepsilon - \varepsilon_0)^2 + \Gamma^2/4} \tag{46}$$

as

$$\begin{aligned}S^<(\omega) = \int \frac{d\varepsilon}{2\pi} &\left[4T(\varepsilon)\left(1 - T(\varepsilon+\omega)\right)\sum_{\alpha\neq\beta} f_\alpha(\varepsilon+\omega)\left(1 - f_\beta(\varepsilon)\right) + 4T(\varepsilon)T(\varepsilon+\omega)\sum_\alpha f_\alpha(\varepsilon+\omega)\left(1 - f_\alpha(\varepsilon)\right)\right. \\ &\left.+ \frac{1}{\Gamma_L\Gamma_R} T(\varepsilon)T(\varepsilon+\omega) \sum_{\alpha\beta} f_\alpha(\varepsilon+\omega)\left(1 - f_\beta(\varepsilon)\right)\left(\delta_{\alpha\beta}\frac{\Gamma_\alpha}{\Gamma_{\gamma\neq\alpha}}\omega^2 - (1-\delta_{\alpha\beta})(4\varepsilon\omega + 3\omega^2)\right)\right].\end{aligned}$$
(47)

In the limit $\Gamma \gg eV, \omega$, the terms in the second line go to zero. Thus, with the assumption $\Gamma \gg eV, \omega \gg k_B T$ the noise expression simplifies to

$$\begin{aligned}S^<(\omega) &\approx 4\int \frac{d\varepsilon}{2\pi}\left[T(1-T)\sum_{\alpha\neq\beta} f_\alpha(\varepsilon+\omega)(1-f_\beta(\varepsilon)) + T^2\sum_\alpha f_\alpha(\varepsilon+\omega)(1-f_\alpha(\varepsilon))\right] \\ &= \frac{4}{2\pi}\left[T(1-T)\left[F(\omega+eV) + F(\omega-eV)\right] + 2T^2 F(\omega)\right]\end{aligned}$$
(48)

where $T = 4\Gamma_L\Gamma_R/\Gamma^2$ is the transmission coefficient and $F(x) = x n_B(x)$. This is in agreement with previous works[5,6]. The factor of 4 in front of the parenthesis originates from the fact that the total noise is considered here.

---

[1] D. J. Scalapino, S. R. White, and S. Zhang, Phys. Rev. B **47**, 7995 (1993).
[2] G. D. Mahan, *Many-particle Physics* (Springer, 2010), 3rd ed.
[3] J. Rammer and H. Smith, Rev. Mod. Phys. **58**, 323 (1986).
[4] H. Haug and A.-P. Jauho, *Quantum Kinetics in Transport and Optics of Semiconductors* (Springer, Berlin, 1998).
[5] R. Aguado and L. P. Kouwenhoven, Phys. Rev. Lett. **84**, 1986 (2000).
[6] Y. M. Blanter and M. Büttiker, Phys. Rep. **336**, 1 (2000).



\end{document}